%% file: main.tex
\documentclass[preprint,5p,twocolumn]{elsarticle}

\usepackage{amsmath}
\usepackage{amssymb}
\usepackage{graphicx}
\usepackage{subcaption}
\usepackage{tikz} 
\usepackage{svg}
\usepackage{booktabs}
\usepackage{multirow}
\usepackage{siunitx}
\usepackage[version=4]{mhchem}
\usepackage{geometry}
\usepackage{setspace}
\usepackage{subcaption}
\usepackage[
    colorlinks=true,
    linkcolor=blue,
    citecolor=blue,
    urlcolor=blue
]{hyperref}

\begin{document}
\begin{frontmatter}

% \title{Validation of Multi-Dimensional, Multi-Material Hydrogen Isotope Transport Modeling for Permeability Inference in FLiBe Systems}
\title{Quantifying Multidimensional Transport Effects on Permeability Inference in FLiBe Systems Using a Validation-Informed Modeling Framework}

\author[1]{Huihua Yang\corref{cor1}}
\cortext[cor1]{Corresponding author}
\ead{huihuay@mit.edu}
\author[1]{Abhishek Saraswat}
\author[2]{Weiyue Zhou}
\author[1]{Kevin Woller}
% \author[1]{LIBRA Team}
\author[1]{James Dark}
\author[1]{Chirag Khurana}
\author[1]{Kaelyn Dunnell}
\author[1]{Ethan Peterson}
\author[1]{Remi Delaporte-Mathurin}

\address[1]{Plasma Science and Fusion Center, Massachusetts Institute of Technology, Cambridge, MA 02139, USA}
\address[2]{Department of Nuclear Science and Engineering, Massachusetts Institute of Technology, Cambridge, MA 02139, USA}

\begin{abstract}
Permeability of hydrogen isotopes in molten salts is commonly inferred from permeation experiments using simplified one-dimensional interpretations, which may not capture the coupled transport pathways present in realistic systems. In this work, a multi-dimensional, multi-material hydrogen isotope transport modeling framework implemented in FESTIM is benchmarked against permeation measurements from the HYPERION experiment conducted at the MIT Plasma Science and Fusion Center. The model explicitly resolves transport across molten salt and nickel structures, as well as external boundary conditions, enabling system-level interpretation of the measured permeation fluxes over the temperature range \SIrange{773}{973}{\kelvin}. Rather than relying on idealized one-dimensional formulations for permeability estimation, this study employs a validation-informed inverse framework to assess how multidomain transport and external boundary assumptions influence the permeability inferred from experimental fluxes. Two limiting external boundary conditions, representing ideal coating and uncoated vessel behavior, are used to define a physically motivated envelope for hydrogen isotope exchange with the environment. The model captures the observed magnitude and temperature dependence of permeation fluxes under both conditions, while revealing significant lateral transport and sidewall leakage pathways that are not represented in one-dimensional interpretations. The inferred FLiBe permeability exhibits consistent Arrhenius behavior but spans a range that depends strongly on the assumed boundary conditions, demonstrating that using one-dimensional formulations to describe a permeation experiment may not be adequate to extract accurate permeability. These results provide a physically grounded framework for interpreting permeation measurements in coupled liquid-metal systems and highlight the importance of multidomain transport modeling for reliable property inference in fusion-relevant molten-salt environments.
\end{abstract}


\end{frontmatter}

% ---------- Sections ----------
\input{introduction}
\input{methodology}
\input{results}
\input{discussion}
\input{conclusions}

% ---------- References ----------
\bibliographystyle{unsrt}
\bibliography{refs}

\end{document}

%% file: introduction.tex
\section{Introduction}

Achieving tritium self-sufficiency is a central requirement for sustaining deuterium-tritium (D-T) fusion energy systems, as tritium is scarce in nature \cite{Kovari2017TritiumResources} and must be bred within the reactor blanket to support long-term operation \cite{DelaporteMathurin2025BABY}. Liquid blanket concepts based on lithium-containing materials have therefore been actively pursued, owing to their potential to simultaneously breed tritium, remove heat, and simplify blanket architectures \cite{IHLI2008912}. Among the candidate materials for liquid blanket applications, molten salts such as FLiBe have drawn continued interest because of their intrinsic tritium breeding capability through lithium-based nuclear reactions \cite{Ferry2022LIBRA}, favorable neutronic performance \cite{Sawan2017MSBlanket}, wide liquid operating temperature range, thermophysical stability, and reduced magnetohydrodynamic constraints compared with liquid metals \cite{TAKEUCHI20081082}. These attributes, together with chemical stability and actively developed approaches for tritium extraction \cite{Satoshi2001FlibeRecovery, Fukada2002MassTransport, Fukada2007Recovery} and control \cite{Fukada2006Control,Abe2008JUPITER}, have made FLiBe a leading candidate for molten-salt blanket concepts. Representative examples include the ARC reactor concept, which employs an all-liquid FLiBe blanket \cite{SORBOM2015378}.

% At the MIT Plasma Science and Fusion Center (PSFC), molten-salt blanket concepts are being actively investigated through the LIBRA (Liquid Immersion Blanket: Robust Accountancy) program \cite{Ferry2022LIBRA}, which seeks to demonstrate controlled and scalable tritium breeding in molten salts under fusion-relevant neutron irradiation. This effort follows a staged experimental strategy beginning with the BABY (Build A Better Yield blanket) experiments \cite{DelaporteMathurin2025BABY}, designed to investigate tritium generation, release, and transport in molten-salt breeder systems at progressively larger scales. Initial BABY experiments at the \SI{0.1}{L} scale provide controlled benchmarks for tritium production and transport in FLiBe \cite{DelaporteMathurin2025BABY}, while the ongoing \si{1}{L} BABY experiments extend these studies toward more reactor-relevant conditions and improved tritium accountancy \cite{DelaporteMathurin2026BABY1L}. Together, these experimental efforts provide critical datasets for understanding hydrogen isotope behavior in coupled liquid-metal environments and for developing predictive models of tritium transport in molten-salt blankets.

Hydrogen isotopes in molten FLiBe blanket system undergo coupled transport processes across gas, liquid, and solid domains, including generation through lithium breeding reactions, dissolution into molten salt, diffusion within the liquid, thermodynamic partitioning at liquid-solid interfaces, and transport through structural materials. Accurate prediction of these processes requires not only reliable transport models, but also well-characterized material properties, especially diffusivity and solubility (permeability), which governs the overall flux of hydrogen isotopes through coupled domains.

Despite its importance, reported hydrogen isotope permeability data for molten FLiBe remain scarce \cite{Saraswat2026FLiBePermeation} and often exhibit variability spanning approximately one to two orders of magnitude across the literature\cite{Nakamura2015FluorideSalt,Nishiumi2016FLINABE,Anderl2004Flibe,Calderoni2008FLIBE}. This variability is commonly attributed to material-specific factors such as salt composition, redox state, and impurity levels \cite{Lam2021HydrogenValence}, as well as differences in experimental design and measurement conditions \cite{Tijssen2025Permeation}. However, an additional and often overlooked source of discrepancy arises from the interpretation of permeation experiments themselves. Most permeability measurements are analyzed using simplified one-dimensional (1D) transport models, in which transport is assumed to occur along a single effective pathway within a single material under idealized boundary conditions. 

In realistic experimental systems, however, hydrogen isotope transport occurs through multiple coupled pathways. In addition to the primary permeation route across a liquid-solid interface (e.g., a molten salt-membrane interface), hydrogen isotope may diffuse through surrounding structural materials, migrate along sidewalls, and exchange with external environments. These multidomain transport processes can introduce parallel pathways and leakage mechanisms that alter the effective transport resistance of the system. As a result, the measured permeation flux reflects a coupled system response rather than transport through a single material along a single dimension.

In this context, the limitation of conventional 1D models is not simply geometric, but fundamental to their formulation: they cannot represent competing and parallel transport pathways. Consequently, permeability values \textit{inferred} from such models may not correspond to \textit{intrinsic} material properties, but instead correspond to system-dependent parameters that embed the effects of geometry, boundary conditions, and multidimensional transport. This raises a key question: to what extent do multidomain transport pathways and external boundary conditions influence the permeability inferred from permeation experiments, and can these effects be quantified in a physically consistent framework?

To address this question, we consider hydrogen isotope permeation measurements from the HYPERION (HYdrogen PERmeatION) facility~\cite{Cota2024HYPERION} at the MIT PSFC, which provides a systematically controlled dataset for hydrogen isotope permeation in coupled gas-liquid-solid systems relevant to FLiBe-based blanket concepts. The experimental configuration consists of a molten-salt region, a nickel membrane, and a surrounding nickel containment structure. While the salt-to-membrane route represents the intended permeation pathway, transport through the structural sidewalls was experimentally identified as a significant contributor to the measured flux. These experiments further reveal key sources of variability in the measured permeation response, including interfacial bubble formation and radial transport losses through the vessel structure~\cite{Saraswat2026FLiBePermeation}. The coupled nature of these transport pathways, together with the inherent limitations of experimental observability, makes it difficult to isolate intrinsic material properties from the measured flux alone, motivating the need for a physically consistent, multidimensional modeling framework.

To this end, a multidimensional, multi-material transport framework implemented in FESTIM \cite{Dark2024FESTIM2, DELAPORTEMATHURIN2024786} is employed to this system to explicitly resolve hydrogen isotope transport across molten salt, metallic domains, and external boundaries. The model is benchmarked against measured permeation fluxes and used within a validation-informed inverse framework to infer the FLiBe permeability, such that the extracted value reflects the salt transport response rather than a convolution of material and geometric effects. 

The present work examines how multidimensional transport pathways and hydrogen isotope exchange with the environment at the vessel boundary affect the interpretation of permeation measurements. The multidimensional pathways are captured directly by the model geometry, while the boundary exchange poses an additional difficulty: the effective behavior of the external vessel surface depends on the uncertain performance of protective coatings at elevated temperatures. Two limiting boundary conditions, corresponding to ideal coating and uncoated vessel behavior, are considered to bracket the possible interaction between the system and its environment. This approach provides a structured basis for assessing the sensitivity of inferred transport behavior to modeling assumptions and experimental conditions.

The main contribution of this work is to demonstrate that multidimensional transport and external boundary assumptions can significantly affect permeability inference in molten-salt systems, and to provide a physically grounded framework for interpreting permeation experiments under realistic, coupled transport conditions.  

The remainder of this paper is organized as follows. Section \ref{sec:methodology} describes the HYPERION experimental configuration and the multidimensional transport modeling framework, and presents the methodology used for permeability inference, including the treatment of experimental and modeling uncertainties. Section \ref{sec:results} presents the FESTIM simulation results for the HYPERION system and the inferred FLiBe permeability under different boundary conditions assumptions. Section \ref{sec:discussions} discusses the results, including the role of structural transport pathways and the comparison between one-dimensional and multidimensional models. Finally, Section \ref{sec:conclusions} summarizes the main findings and outlines implications for permeability interpretation in molten-salt systems.

%% file: methodology.tex
\section{Methodology}
\label{sec:methodology}
\subsection{Hyperion Experimental Configuration}

\begin{figure}[htbp]
    \centering
    \includegraphics[width=\linewidth]{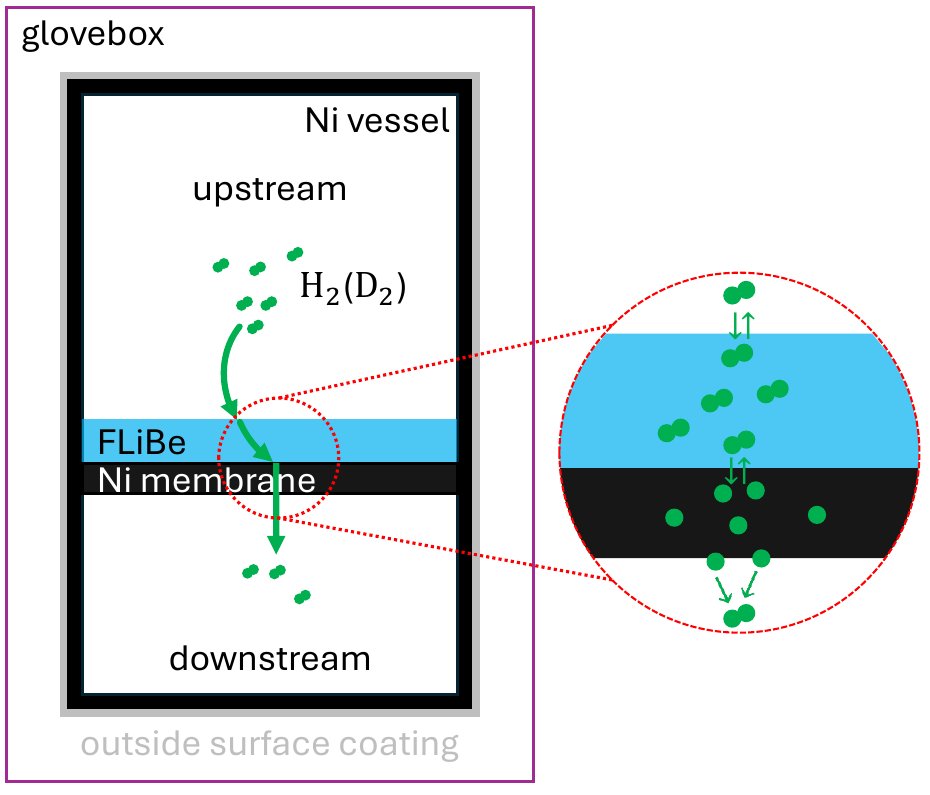}
    \caption{Experimental configuration of the HYPERION system.}
    \label{fig:hyperion_framework}
\end{figure}

The Hyperion experiment, conducted at the MIT PSFC, is designed to investigate hydrogen isotope permeation in coupled gas-molten salt-metal systems representative of FLiBe-based blanket environments. A detailed description of the experimental design and operating procedures is provided in our previous work \cite{Saraswat2026FLiBePermeation}. Traditional interpretations of liquid permeation experiments rely on one-dimensional formulations. In previous HYPERION studies, this approach was extended with lumped correction terms to account for sidewall losses \cite{Saraswat2026FLiBePermeation}. The present work extends this approach by directly resolving multidimensional transport processes, enabling a more detailed interpretation of the coupled salt-metal system. The configuration considered in this study is schematically illustrated in Fig.~\ref{fig:hyperion_framework}. A nickel (Ni-200) membrane is integrated within a nickel containment structure, separating an upstream molten-salt region from a downstream gas measurement region.

Hydrogen (\ce{H2}) or deuterium (\ce{D2}) gas is introduced on the molten-salt side under controlled temperature and pressure conditions. Molecular gas establishes an equilibrium partial pressure at the gas-salt interface and dissolves in the molten FLiBe. The dissolved species then diffuses through the salt layer toward the nickel membrane. At the salt-metal interface, hydrogen or deuterium partitions according to thermodynamic equilibrium (Section. \ref{sec:interface_condition}) and subsequently diffuses through the nickel membrane.

The downstream side of the membrane is continuously purged with an inert carrier gas (\ce{Ar}) at a prescribed volumetric flow rate. Hydrogen or deuterium permeating through the membrane is entrained in the sweep stream and transported to a gas chromatography (GC) system for analysis. 

In addition to the intended permeation pathway through the molten salt and membrane, experimental observations indicate the presence of additional transport pathways associated with the containment structure. In particular, hydrogen accumulation has been detected in the glovebox atmosphere during operation, necessitating periodic purging of the glovebox to maintain a certain range of background gas concentration. This observation suggests that hydrogen or deuterium escapes through the vessel walls or other structural components into the surrounding environment.

\begin{figure}[htbp]
    \centering
    \includegraphics[width=\linewidth]{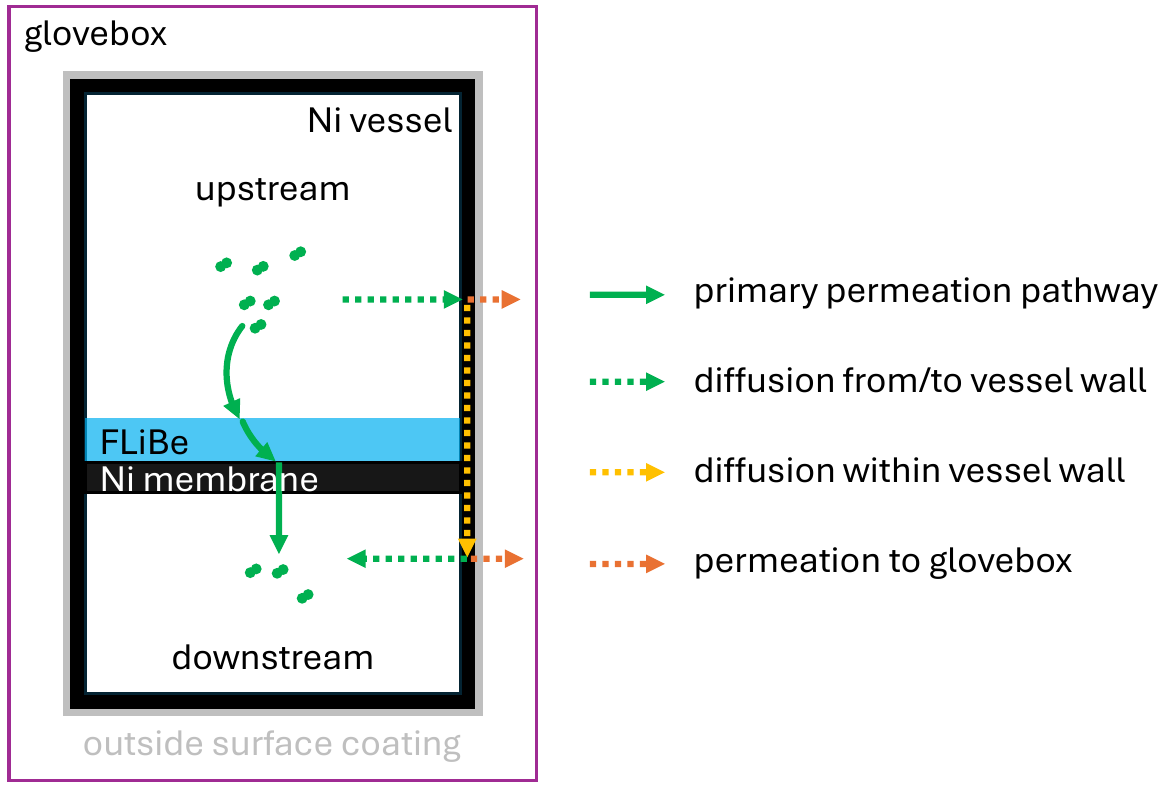}
    \caption{Possible transport pathways in the HYPERION system.}
    \label{fig:flow_path}
\end{figure}

Such structural transport may take two distinct forms. First, species dissolved in the upstream region may diffuse radially through the vessel wall and escape to the glovebox environment, consistent with the observed glovebox accumulation. Second, diffusion within the structural material may create bypass pathways connecting the upstream and downstream regions outside the direct membrane route, contributing to the measured permeation signal. A schematic representation of these possible transport pathways is illustrated in Fig.~\ref{fig:flow_path}.

The geometric configuration therefore gives rise to a multidimensional transport problem rather than an idealized one-dimensional membrane diffusion scenario. Reliable interpretation of the measured fluxes therefore requires a modeling framework that resolves the coupled transport across gas, liquid, and solid domains.

\subsection{Flux determination and measurement uncertainty}
\label{sec:UQ}
The total permeation flux is determined from the measured outlet mole fraction together with the known sweep-gas flow rate and thermodynamic state of the gas stream. Assuming ideal-gas behavior, the hydrogen isotope permeation flux is expressed as:

\begin{equation}
J = N_\mathrm{A} \cdot Q \frac{yP}{RT},
\label{eq:flux-def}
\end{equation}
where $J$ is the GC-measured permeation flux of H or D atoms (\si{atoms\per\second}), $Q$ is the volumetric sweep-gas flow rate (\si{m^{3}.s^{-1}}) evaluated at pressure $P$ (\si{\pascal}) and temperature $T$ (\si{\kelvin}), $y$ is the hydrogen isotope mole fraction in the sweep gas measured by gas chromatography (GC), $N_\mathrm{A} = \SI{6.022e23}{mol^{-1}}$ is Avogadro's number, and $R$ is the universal gas constant (8.314 \si{ J.mol^{-1}.K^{-1}}).

\begin{figure}[htbp]
    \centering
    \includegraphics[width=\linewidth]{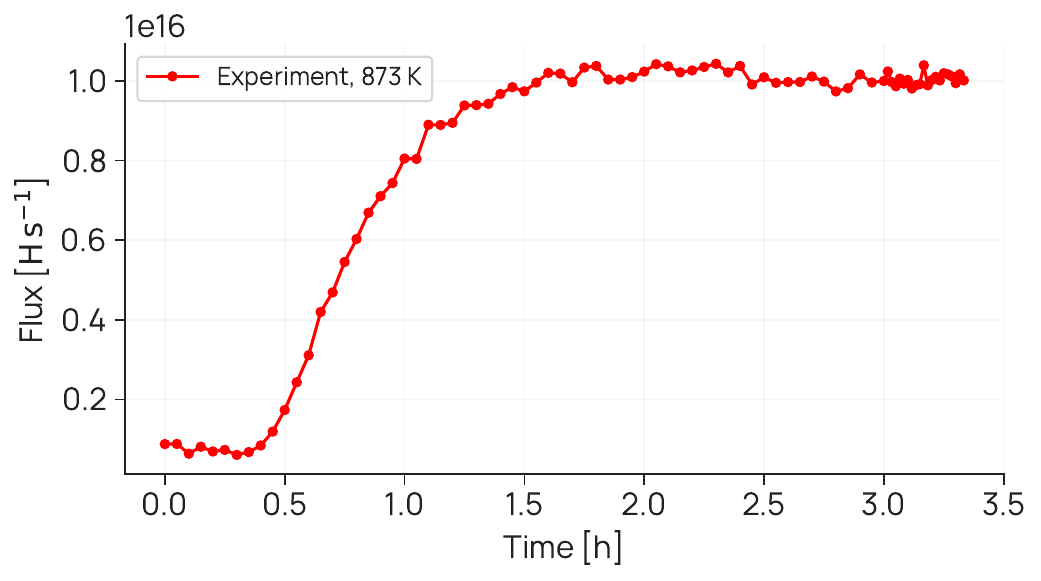}
    \caption{Time-dependent experimentally measured permeation flux by GC at $T =$ \SI{873}{\kelvin}.}
    \label{fig:experimental_data}
\end{figure}

An example of the time-dependent experimental permeation flux at $T =$ \SI{873}{\kelvin} is shown in Fig.~\ref{fig:experimental_data}.

While the full time-dependent signal is recorded, only the steady-state portion is analyzed in this work. For each experimental condition, multiple steady-state GC measurements are recorded. The representative mole fraction is obtained by averaging $M$ steady-state samples,

\begin{equation}
\bar{y}=\frac{1}{M}\sum_{i=1}^{M}y_i,
\end{equation}

and the corresponding steady-state permeation flux measured by GC is

\begin{equation}
\bar{J}=N_\mathrm{A} \cdot Q \frac{\bar{y}P}{RT}.
\end{equation}

Uncertainty in the flux arises from both measurement repeatability and instrumentation specifications. Following the ISO Guide to the Expression of Uncertainty in Measurement \cite{ISO_GUM2008}, uncertainties are classified as Type A (statistical) and Type B (instrumental) contributions.

The repeatability of GC measurements is quantified using the sample standard deviation,

\begin{equation}
s_y=\sqrt{\frac{1}{M-1}\sum_{i=1}^{M}(y_i-\bar{y})^2},
\end{equation}
which yields the Type A standard uncertainty in the averaged mole fraction,

\begin{equation}
u_A(\bar{y})=\frac{s_y}{\sqrt{M}}.
\end{equation}

Because the permeation flux is linear in $y$, the corresponding Type A uncertainty in flux is

\begin{equation}
u_A(\bar{J})=\bar{J}\frac{u_A(\bar{y})}{\bar{y}}.
\end{equation}

Additional uncertainty arises from instrumentation specifications associated with the sweep-gas flow rate, GC measurement accuracy and pressure, as summarized in Table~\ref{tab:uncertainty}.

\begin{table}[htbp]
\centering
\caption{Instrumentation uncertainties used in flux uncertainty propagation.}
\label{tab:uncertainty}
\begin{tabular}{p{4cm}r}
\hline
Quantity & Relative uncertainty \\
\hline
Sweep-gas flow rate ($Q$) & \SI{3}{\percent} \\
GC mole fraction ($y$) & \SI{1}{\percent} \\
Pressure ($P$) & \SI{1}{\percent} \\
\hline
\end{tabular}
\end{table}

These contributions are treated as Type B uncertainties and combined using first-order uncertainty propagation,

\begin{equation}
\begin{aligned}
\left(\frac{u_B(\bar{J})}{\bar{J}}\right)^2=
\left(\frac{u_B(Q)}{Q}\right)^2 +
\left(\frac{u_B(\bar{y})}{\bar{y}}\right)^2 +
\left(\frac{u_B(P)}{P}\right)^2
\end{aligned}
\end{equation}

The combined standard uncertainty in the steady-state flux is then obtained as

\begin{equation}
u(\bar{J})=\sqrt{u_A^2(\bar{J})+u_B^2(\bar{J})}.
\end{equation}

While the above treatment quantifies measurement-driven uncertainty in the experimentally flux, additional sources of uncertainty associated with model assumptions (e.g., nickel permeability correlations, external boundary conditions, and temperature-dependent FLiBe geometry) are addressed separately within the permeability inference framework (section \ref{sec:perme_infer}).

\subsection{FESTIM Multi-Dimensional, Multi-Material Transport Modeling Framework}

\begin{figure}[h!]
    \centering
    \includegraphics[width=\linewidth]{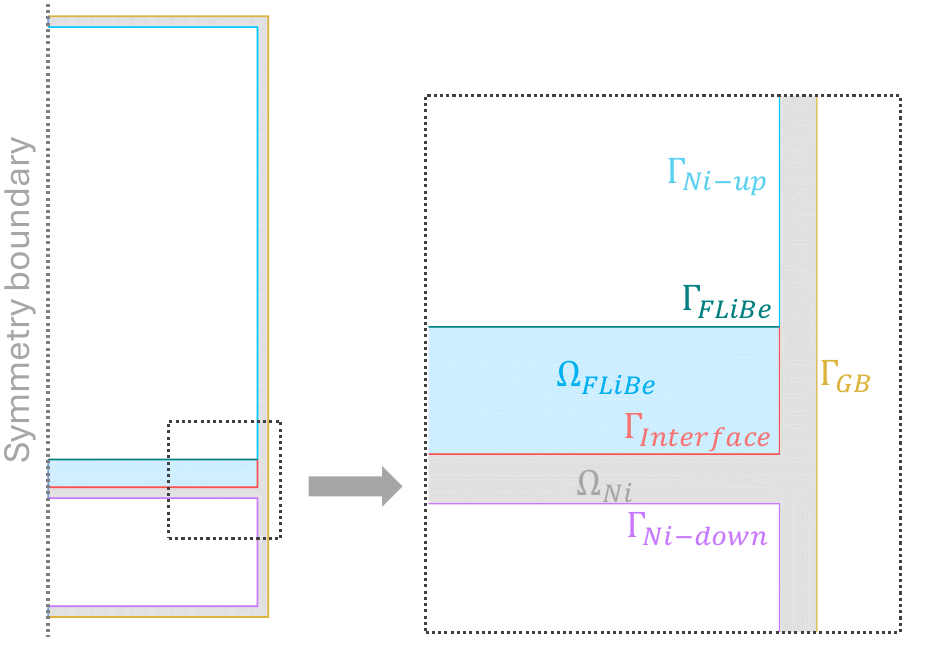}
    \caption{Schematic of the computational domain used in the HYPERION model. The left panel shows the half-geometry with an axial symmetry. The right panel presents a zoomed-in view of the region near the liquid-solid interface.}
    \label{fig:mesh_hyperion}
\end{figure}

Hydrogen isotope transport in the HYPERION configuration is formulated using a unified multi-domain framework applicable to both hydrogen (H) and deuterium (D). The cylindrical geometry of the HYPERION cell is modeled directly in FESTIM using its axisymmetric (r-z) capability, which reduces a three-dimensional transport problem to a two-dimensional formulation without loss of physical fidelity. The computational domain, shown in Fig.~\ref{fig:mesh_hyperion}, consists of coupled molten FLiBe and nickel regions, denoted by $\Omega_i$, where the subscript $i$ identifies the material domain (e.g., FLiBe or Ni). A symmetry condition is imposed along the central axis.

In this formulation, isotope effects are incorporated through isotope-specific variables and material properties. The concentration $c_{i,k}(\mathbf{x},t)$ denotes the concentration of isotope $k \in \{\mathrm{H}, \mathrm{D}\}$ within domain $\Omega_i$. Accordingly, all transport and thermodynamic parameters are defined consistently for each isotope.

Hydrogen isotopes are introduced at the upstream gas-salt interface and subsequently undergo a sequence of coupled transport processes: dissolution into molten FLiBe, diffusion within the salt domain $\Omega_{\mathrm{FLiBe}}$, thermodynamic partitioning across the salt-metal interface $\Gamma_{\mathrm{Interface}}$, and diffusion through the nickel domain $\Omega_{\mathrm{Ni}}$. On the downstream side, isotopes permeate into a sweep gas, while additional transport may occur through the surrounding nickel structure toward the glovebox environment via the external boundary $\Gamma_{\mathrm{GB}}$.

The present study focuses on steady-state conditions, under which the measured permeation flux $J_k$ reflects a global balance among bulk diffusion, interfacial thermodynamic partitioning, and boundary-driven exchange processes.

\subsubsection{Governing Equations}

Within each material domain $\Omega_i$, the concentration $c_{i,k}$ of isotope $k$ satisfies the diffusion equation:

\begin{equation}
\frac{\partial c_{i,k}}{\partial t}
=
\nabla \cdot \left( D_{i,k} \nabla c_{i,k} \right),i=\{\Omega_{\mathrm{Ni}}, \Omega_{\mathrm{FLiBe}}\}, k=\{H, D\},
\end{equation}
where $D_{i,k}$ represents the diffusivity of isotope $k$ in material $i$. It is expressed using an Arrhenius relation,

\begin{equation}
% D_{i,k} = D_{0,i,k} \exp\left(-\frac{E_{D,i,k}}{RT}\right).
D_{i,k} = D_{0,i,k} \exp\left(-\frac{E_{D,i,k}}{k_\mathrm{B}T}\right)
\end{equation}
where $D_{0,i,k}(\si{m^2\,s^{-1}})$ is the pre-exponential diffusivity factor of isotope $k$ in material $i$, 
%$E_{D,i,k}(\si{J\,mol^{-1}})$ is the corresponding activation energy. 
$E_{D,i,k}\,(\si{eV})$ is the corresponding activation energy, and $k_\mathrm{B} = \SI{8.617e-5}{eV\,K^{-1}}$ is the 
Boltzmann constant.

\subsubsection{Interface Condition}
\label{sec:interface_condition}

Hydrogen isotope solubility differs between molten FLiBe and nickel. In molten FLiBe, isotopes obey Henry's law of solubility. The effective partial pressure $P_k$ (\si{\pascal}) of hydrogen isotope $k \in \{\mathrm{H}, \mathrm{D}\}$ in thermodynamic equilibrium with the dissolved species is given by \cite{DELAPORTEMATHURIN2024786}:

\begin{equation}
P_k = \frac{c_{\mathrm{FLiBe},k}}{K_{H,k}}.
\end{equation}

In nickel, hydrogen isotopes obey Sieverts' law of solubility, and the corresponding equilibrium partial pressure is expressed as:

\begin{equation}
P_k  = \left( \frac{c_{\mathrm{Ni},k} }{K_{S,k} } \right)^2.
\end{equation}
where $c_{\mathrm{FLiBe},k}$ and $c_{\mathrm{Ni},k}$ denote the concentrations 
%(mol\,m$^{-3}$) 
(\si{atom\,m^{-3}}) of isotope $k$ in molten FLiBe and nickel, respectively. $K_{H,k}$ 
%(mol\,m$^{-3}$\,Pa$^{-1}$) 
(\si{atom\,m^{-3}\,Pa^{-1}}) is the Henry-type solubility constant in FLiBe, while $K_{S,k}$ 
%(mol\,m$^{-3}$\,Pa$^{-1/2}$) 
(\si{atom\,m^{-3}\,Pa^{-1/2}}) is the Sieverts-type solubility constant in nickel. 

The solubility constants exhibit temperature dependence and are described using Arrhenius-type relations. For molten FLiBe,

\begin{equation}
%K_{H,k} = K_{H,0,k} \exp\left(-\frac{E_{K_H,k}}{RT}\right),
K_{H,k} = K_{H,0,k} \exp\left(-\frac{E_{K_H,k}}{k_\mathrm{B}T}\right)
\end{equation}

and for nickel,

\begin{equation}
%K_{S,k} = K_{S,0,k} \exp\left(-\frac{E_{K_S,k}}{RT}\right),
K_{S,k} = K_{S,0,k} \exp\left(-\frac{E_{K_S,k}}{k_\mathrm{B}T}\right)
\end{equation}
where $K_{H,0,k}$ 
%(\si{mol\,m^{-3}\,Pa^{-1}})
(\si{atom\,m^{-3}\,Pa^{-1}}) and $K_{S,0,k}$
%(\si{mol\,m^{-3}\,Pa^{-1/2}})
(\si{atom\,m^{-3}\,Pa^{-1/2}}) are the pre-exponential solubility constants of isotope $k$ in FLiBe and nickel, respectively. $E_{K_H,k}$ 
%(\si{J\, mol^{-1}}) 
(\si{eV}) and $E_{K_S,k}$
%(\si{J\, mol^{-1}})
(\si{eV}) are the corresponding effective dissolution enthalpies.

At the liquid-solid interface $\Gamma_{\mathrm{Interface}}$ (Fig.~\ref{fig:mesh_hyperion}), local thermodynamic equilibrium is assumed. This implies continuity of chemical potential for each isotope across the interface. For hydrogen isotopes, this condition can be expressed in terms of an equivalent partial pressure, yielding

\begin{equation}
\frac{c_{\mathrm{FLiBe},k}}{K_{H,k}}
=
\left( \frac{c_{\mathrm{Ni},k}}{K_{S,k}} \right)^2,\quad \mathrm{on~\Gamma_{interface}} 
\end{equation}

\subsubsection{Boundary Conditions}

Boundary conditions are applied consistently for both isotopes.

At the upstream boundary, hydrogen charging is imposed through a prescribed partial pressure $P_{\mathrm{up},k}$,

\begin{equation}
c_{\mathrm{FLiBe},k} = K_{H,k} P_{\mathrm{up},k},\quad \mathrm{on~\Gamma_{FLiBe}}, 
\end{equation}

\begin{equation}
c_{\mathrm{Ni-up},k} = K_{S,k} \sqrt{P_{\mathrm{up},k}}, \quad \mathrm{on~\Gamma_{Ni-up}}. 
\end{equation}

At the downstream boundary, equilibrium with the sweep gas yields:

\begin{equation}
c_{\mathrm{Ni-down},k} = K_{S,k} \sqrt{P_{\mathrm{down},k}}, \quad \mathrm{on~\Gamma_{Ni-down}}.
\end{equation}

A symmetry condition is imposed along the left boundary,

\begin{equation}
\nabla c_{k} \cdot \mathbf{n} = 0.
\end{equation}

The external boundary condition will be discussed further in Section \ref{sec:coating_condition}.

\subsection{Permeability inference framework}
\label{sec:perme_infer}

Permeability inference in the HYPERION system is formulated as a nonlinear inverse problem: the FLiBe permeability is adjusted such that the simulated steady-state flux matches the measured flux under specified thermodynamic and boundary conditions. Because the measured response reflects coupled transport across multiple domains and depends on the assumed external boundary behavior, the inferred permeability is a system-consistent parameter conditioned on the assumed geometry, transport model, and boundary representation.

\subsubsection{Inverse formulation}

Under steady-state conditions, the permeation flux across FLiBe is governed by the permeability, defined as
\begin{equation}
\Phi_k = D_k K_{H,k},
\label{eq:phi_definition}
\end{equation}
where $\Phi_k$ is the permeability 
%(\si{mol.m^{-1}.s^{-1}.Pa^{-1}}).
(\si{atom \,\meter^{-1}\,\second^{-1}\,\pascal^{-1}}).

\begin{figure*}[ht]
    \centering
    \includegraphics[width=\linewidth]{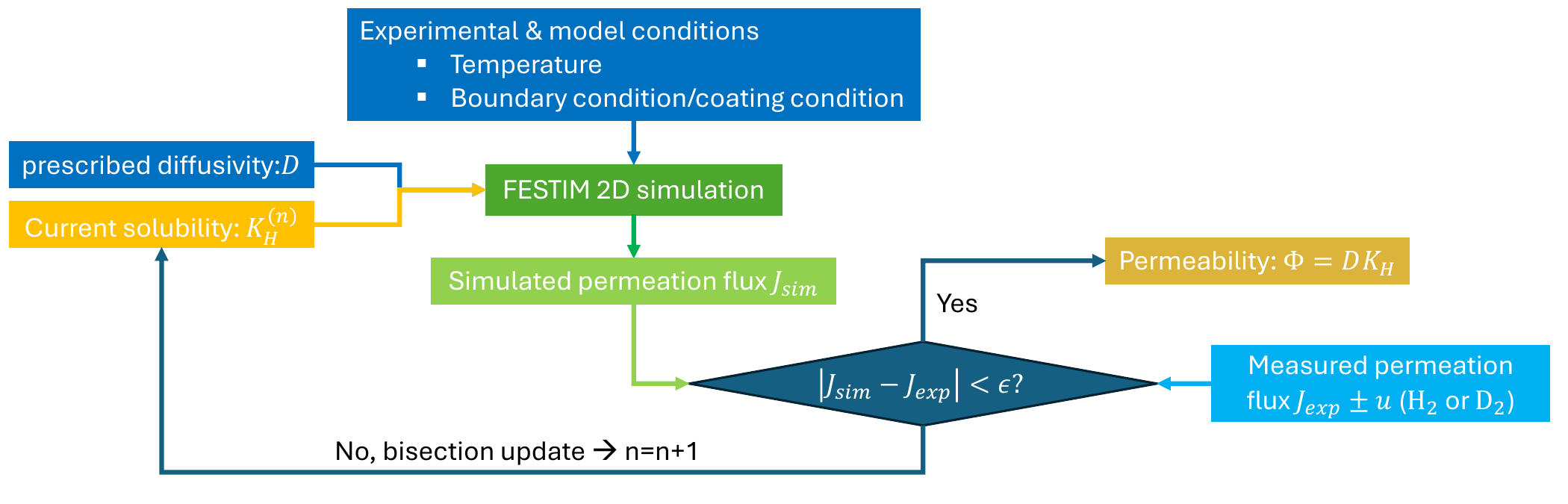}
    \caption{Schematic of the permeability inference framework. A trial solubility is iteratively updated using a bisection algorithm until the simulated permeation flux matches the measured flux within a prescribed tolerance.}
    \label{fig:flowchart}
\end{figure*}

The permeability is inferred through an iterative inverse procedure, as illustrated in Fig.~\ref{fig:flowchart}. the diffusivity $D_k$ is prescribed, and the solubility $K_{H,k}$ is treated as the adjustable parameter in the inverse procedure. For each experimental condition, a trial solubility $K_{H,k}^{(n)}$ is specified and a multidimensional transport simulation is performed using FESTIM to compute the corresponding steady-state permeation flux $J_{\mathrm{sim}}$.

The inverse problem is solved by iteratively updating the solubility through a bisection algorithm until the simulated flux matches the experimentally measured flux $J_{\mathrm{exp}}$ within a prescribed tolerance $\epsilon$:
\begin{equation}
\left| J_{\mathrm{sim}}\left(K_{H,k}^{(n)}\right) - J_{\mathrm{exp}} \right| < \epsilon.
\end{equation}

Once convergence is achieved, the inferred solubility is used to compute the corresponding permeability through Eq.~(\ref{eq:phi_definition}). This procedure yields a permeability value that is consistent with the coupled transport processes represented in the model and the specified experimental conditions.

\subsubsection{External boundary condition envelope}
\label{sec:coating_condition}

The external surface of the nickel containment structure is in contact with the glovebox atmosphere, where hydrogen exchange may occur through outgassing or re-entry. However, the effective mass-transfer behavior at this boundary is uncertain due to the protective coatings (permeation barriers) whose performance at elevated temperatures is not well characterized.

To account for this uncertainty, two limiting boundary conditions are considered for the external surface $\Gamma_{\mathrm{GB}}$ (Fig.~\ref{fig:mesh_hyperion}). In the first limit, an ideal coating is assumed, corresponding to a zero-flux condition that suppresses hydrogen transport across the surface. In the second limit, an uncoated surface is assumed, fully exposed to the glovebox atmosphere, and local thermodynamic equilibrium is imposed through a Sieverts-type condition.

These two cases define a physically motivated envelope for hydrogen exchange with the external environment. All simulations are performed under both assumptions, and the corresponding permeability values are reported as a bounded range rather than a single value.

The ideal coating boundary condition can mathematically be expressed as:
\begin{equation}
- D_{\mathrm{Ni},k} \nabla c_{\si{Ni},k} \cdot \mathbf{n} = 0 
\quad \text{on } \Gamma_{\mathrm{GB}},
\end{equation}
where $\mathbf{n}$ is the outward unit normal vector.

The uncoated boundary condition is:
\begin{equation}
c_{\mathrm{Ni},k} = K_{S,k} \sqrt{P_{\mathrm{gb},k}}
\quad \text{on } \Gamma_{\mathrm{GB}},
\end{equation}
where $P_{\mathrm{gb},k}$ is the partial pressure of species $k$ in the glovebox (\si{\pascal}).

\subsubsection{Metal-side transport constraints}

Hydrogen transport through the nickel membrane and surrounding structure introduces additional uncertainty in permeability inference. To constrain the metal-side contribution, a nickel-only (dry-run) configuration is considered as a reference case, in which molten FLiBe is absent and hydrogen transport occurs solely through the nickel domain.

The measured dry-run fluxes are first compared with multidimensional simulations using representative nickel permeability correlations from the literature. This comparison verifies that the modeling framework reproduces the metal-side response within the reported range of nickel permeabilities. However, literature nickel permeabilities span a range, and selecting a single value would introduce an arbitrary bias into the subsequent coupled simulations. A system-consistent nickel permeability is then inferred for each external boundary condition by matching the simulated steady-state flux to the dry-run measurements. The resulting parameterization is used in the coupled Ni-FLiBe simulations, reducing the degrees of freedom associated with metal-side transport.

\subsubsection{Temperature-dependent geometry and transport}
\label{sec:T_influence}

In the HYPERION configuration, the effective thickness of the molten FLiBe layer depends on temperature through the salt density. Because the permeation flux is set by the combined transport resistance of the coupled domains, any change in salt thickness alters the diffusion path length and shifts the mapping between permeability and the measured flux.

To ensure consistency across experimental conditions, the salt thickness is updated at each temperature from the measured salt mass and the temperature-dependent density. This treatment prevents geometry-induced variations in transport resistance from being absorbed into the inferred material properties.

The salt density is calculated using \cite{Vidrio2022FLiBeDensity}:

\begin{equation}
\rho_{\mathrm{FLiBe}} =2128.9 - 0.424\,T,
\end{equation}
where $\rho_{\mathrm{FLiBe}}$ is expressed in \si{kg.m^{-3}} and $T$ is in \si{\kelvin}. The corresponding salt volume is:

\begin{equation}
V_{\mathrm{salt}} = \frac{m_{\mathrm{salt}}}{\rho_{\mathrm{FLiBe}}(T)},
\end{equation}
where $m_{\mathrm{salt}}$ is the total salt mass (\si{kg}). The effective salt thickness is:

\begin{equation}
h_{\mathrm{salt}} = \frac{V_{\mathrm{salt}}}{A_{\mathrm{salt}}},
\end{equation}
where $A_{\mathrm{salt}}$ is the cross-sectional area of the salt volume, given by

\begin{equation}
A_{\mathrm{salt}}= \frac{\pi d^2}{4},
\end{equation}
where $d$ is the salt diameter (\si{m}). 

\subsubsection{Measurement uncertainty and error propagation}

The permeation flux is determined from measured gas composition and flow conditions, and is subject to both statistical and instrumental uncertainties, as discussed in Section \ref{sec:UQ}. These uncertainties are propagated to the inferred permeability through first-order sensitivity analysis as follows:

\begin{equation}
u(\Phi_i) \approx \frac{u(J_i)}{\left|\frac{\partial J}{\partial \Phi}\right|_i},
\end{equation}
where $u(\Phi_i)$ and $u(J_i)$ is the uncertainties in the inferred permeability and measured flux at experimental condition, and $\frac{\partial J}{\partial \Phi}$ is the local sensitivity of the simulated flux to the assumed permeability.

This formulation ensures that uncertainty in the experimental measurements is consistently reflected in the inferred transport parameters. The resulting permeability uncertainties are subsequently used to define weights in the Arrhenius regression of permeability as a function of temperature. This approach ensures that the fitted transport parameters reflect the relative confidence in each experimental condition and reduces the influence of higher-uncertainty data points on the inferred temperature dependence.

\subsubsection{Arrhenius representation of inferred permeability}

The temperature dependence of the inferred permeability is represented using an Arrhenius relation as follows:
\begin{equation}
\Phi_k(T) = \Phi_{0,k} \exp\left(-\frac{E_{\Phi,k}}{RT}\right),
\end{equation}
where $\Phi_{0,k}$ is the pre-exponential factor 
%(\si{mol-H.m^{-1}.s^{-1}.Pa^{-1}})
(\si{atom\,m^{-1}\,s^{-1}\,Pa^{-1}}) of hydrogen isotope $k$, $E_{\Phi,k}$ is the corresponding activation energy 
%(\si{kJ.mol^{-1}})
(\si{eV}). These parameters are obtained by weighted linear regression of $\ln \Phi_k$ against $\si{1/T}$, with the weights defined in the previous section. The fit is performed independently for each external boundary-condition case, and the resulting Arrhenius parameters are reported as a range bracketing the two limits.

%% file: results.tex
\section{Results}
\label{sec:results}
\subsection{Influence of coating condition}
\label{sec:coating}

The influence of external boundary conditions on hydrogen isotope transport is first examined by comparing simulations performed under ideal coating and uncoated assumptions. In both cases, identical permeability properties are used, and the upstream and downstream pressures are held constant, so that any differences in transport behavior arise solely from the external boundary treatment.

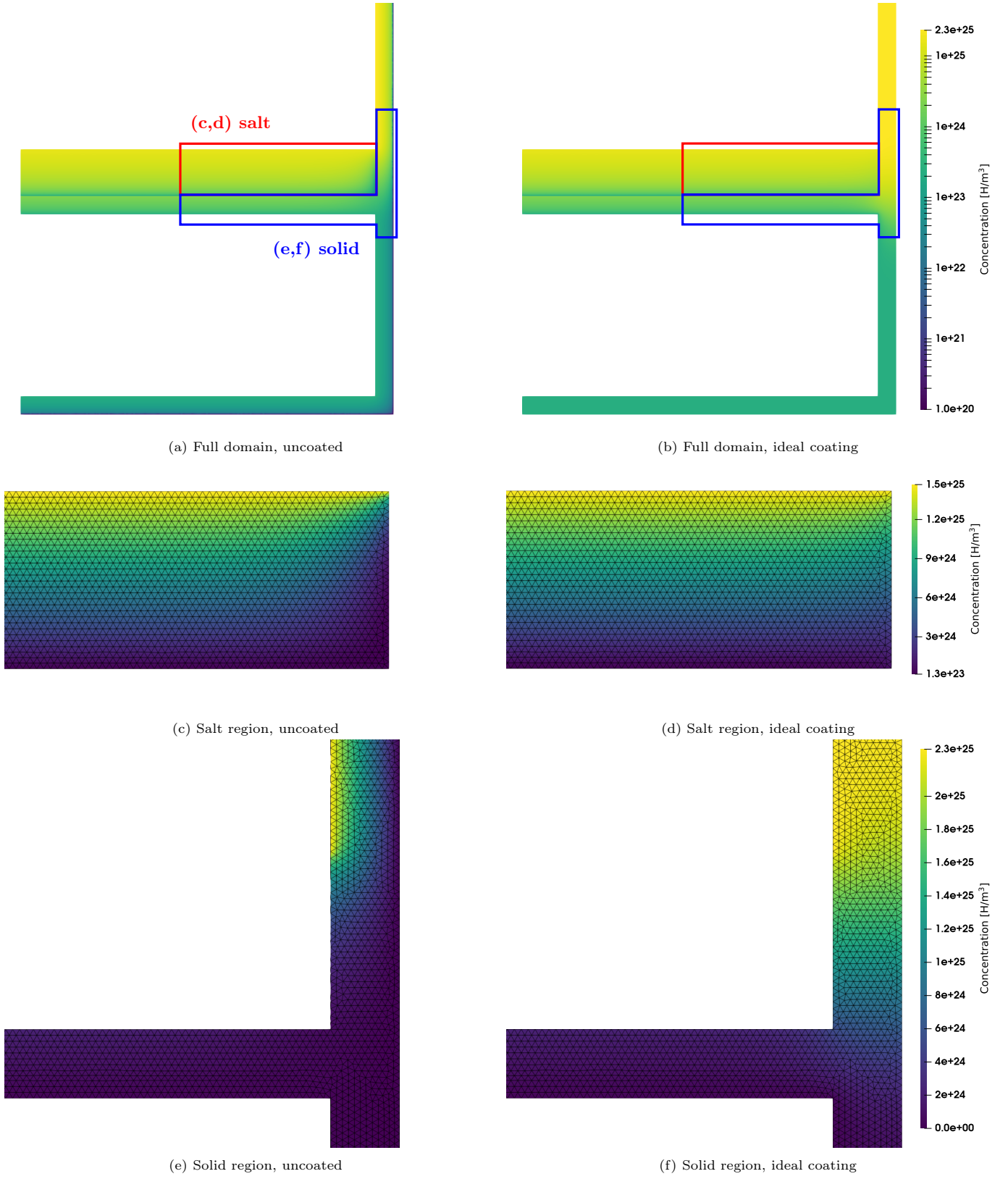
\begin{figure*}[htbp]
    \centering
    %full
    \begin{subfigure}[b]{0.5\linewidth}
        \begin{tikzpicture}
            \node[anchor=south west, inner sep=0] (image) at (0,0) {
                \includegraphics[width=\linewidth,trim=0pt 0pt 0pt 600pt, clip]{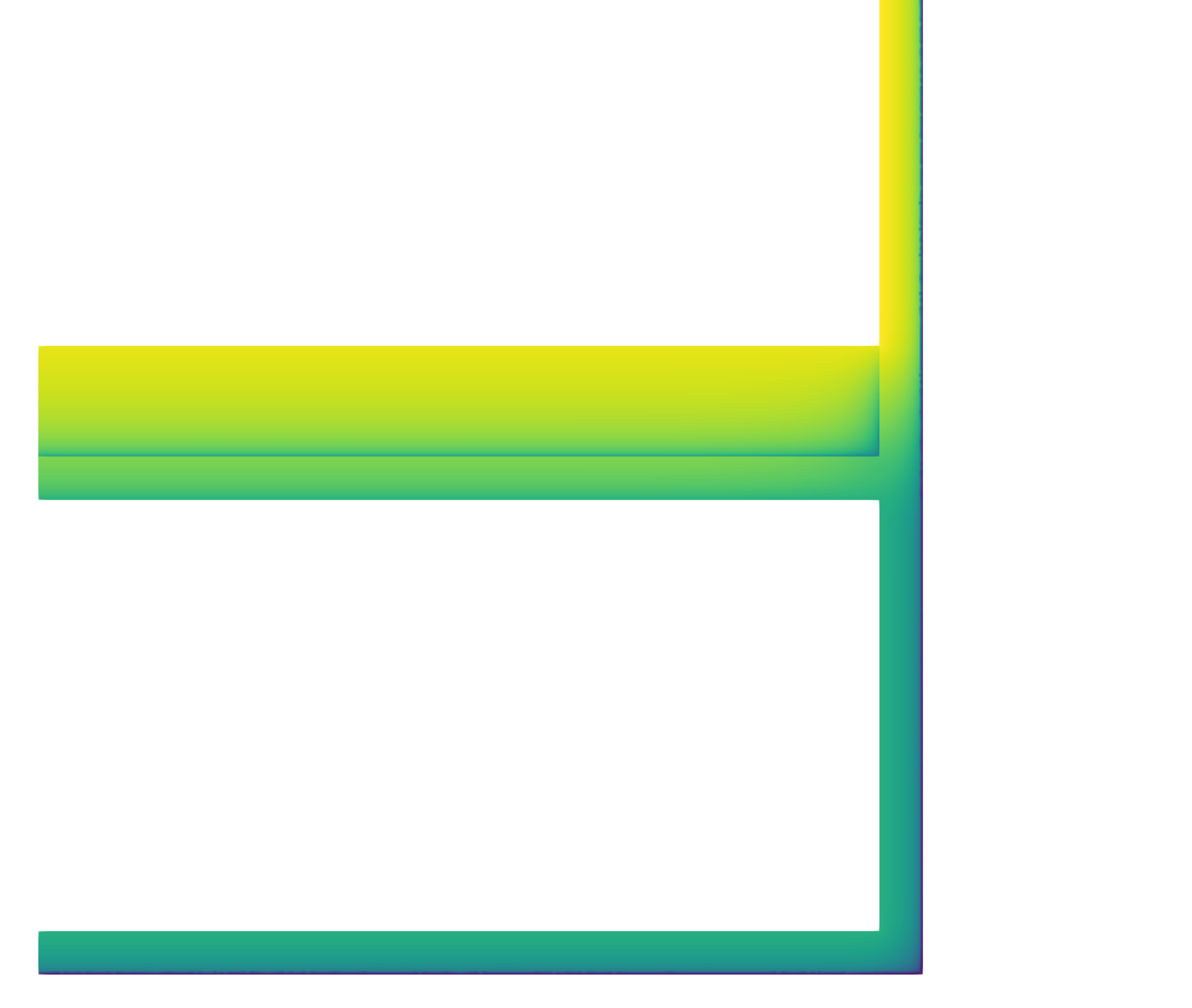}
            };
            \begin{scope}[x={(image.south east)}, y={(image.north west)}]
                \draw[red, very thick] (0.35, 0.55) rectangle (0.74, 0.67);
                \node[red, font=\small\bfseries, above] at (0.45, 0.68) {(c,d) salt};                
                \draw[blue, very thick] 
                    (0.74, 0.75) -- (0.78, 0.75) -- (0.78, 0.45) -- 
                    (0.74, 0.45) -- (0.74, 0.48) -- (0.35, 0.48) -- 
                    (0.35, 0.55) -- (0.74, 0.55) -- cycle;
                
                \node[blue, font=\small\bfseries, right] at (0.52, 0.42) {(e,f) solid};
            \end{scope}
        \end{tikzpicture}
        \caption{Full domain, uncoated}
        \label{fig:uncoated_full}
    \end{subfigure}%
    \begin{subfigure}[b]{0.5\linewidth}
        \begin{tikzpicture}
            \node[anchor=south west, inner sep=0] (image) at (0,0) {
                \includegraphics[width=\linewidth,trim=0pt 0pt 0pt 600pt, clip]{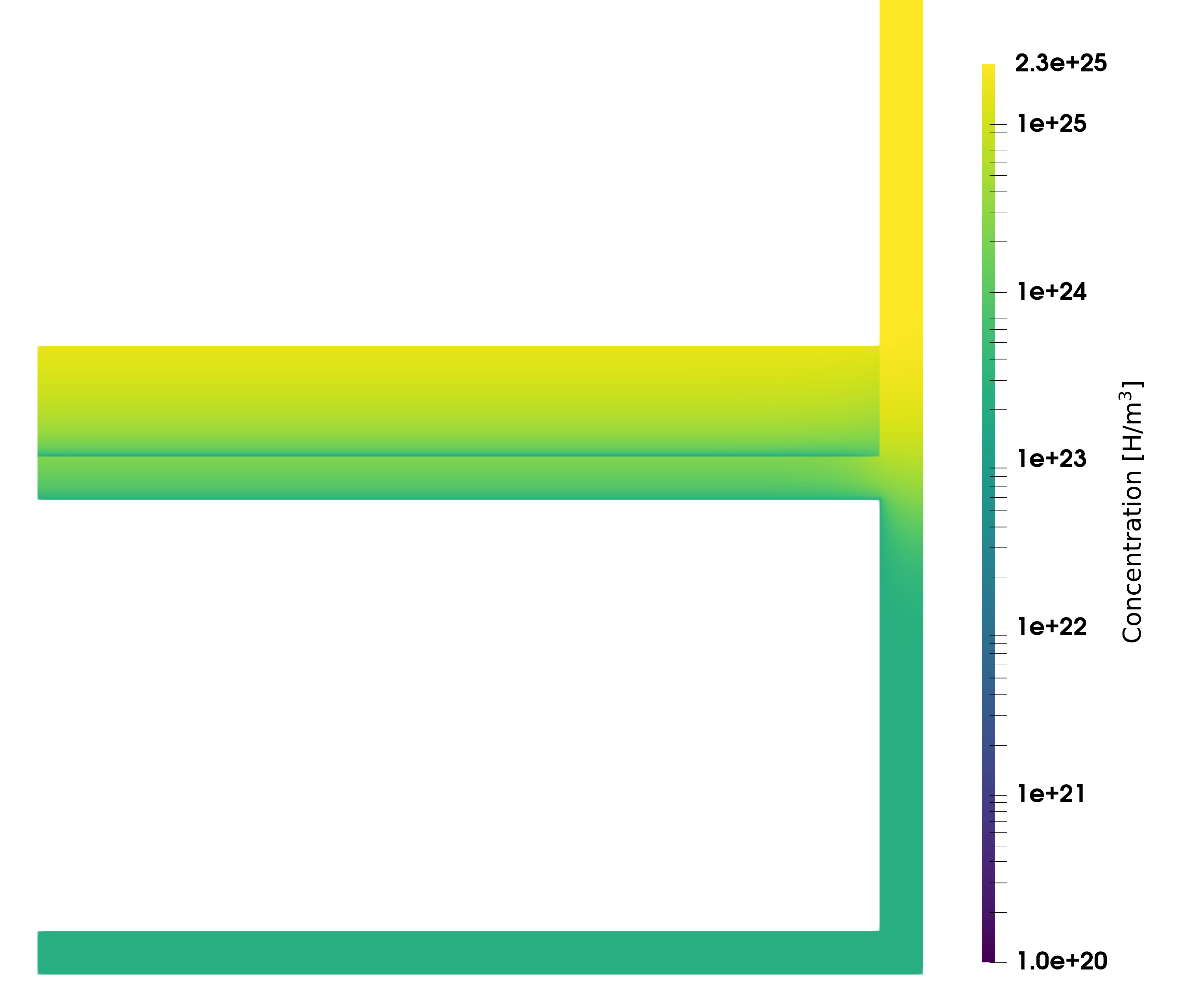}
            };
\begin{scope}[x={(image.south east)}, y={(image.north west)}]
                \draw[red, very thick] (0.35, 0.55) rectangle (0.74, 0.67);              
                \draw[blue, very thick] 
                    (0.74, 0.75) -- (0.78, 0.75) -- (0.78, 0.45) -- 
                    (0.74, 0.45) -- (0.74, 0.48) -- (0.35, 0.48) -- 
                    (0.35, 0.55) -- (0.74, 0.55) -- cycle;
            \end{scope}
        \end{tikzpicture}
        \caption{Full domain, ideal coating}
        \label{fig:ideal_full}
    \end{subfigure}

    % salt
    \begin{subfigure}[b]{0.5\linewidth}
        \includegraphics[width=\linewidth,trim=0pt 650pt 0pt 450pt, clip]{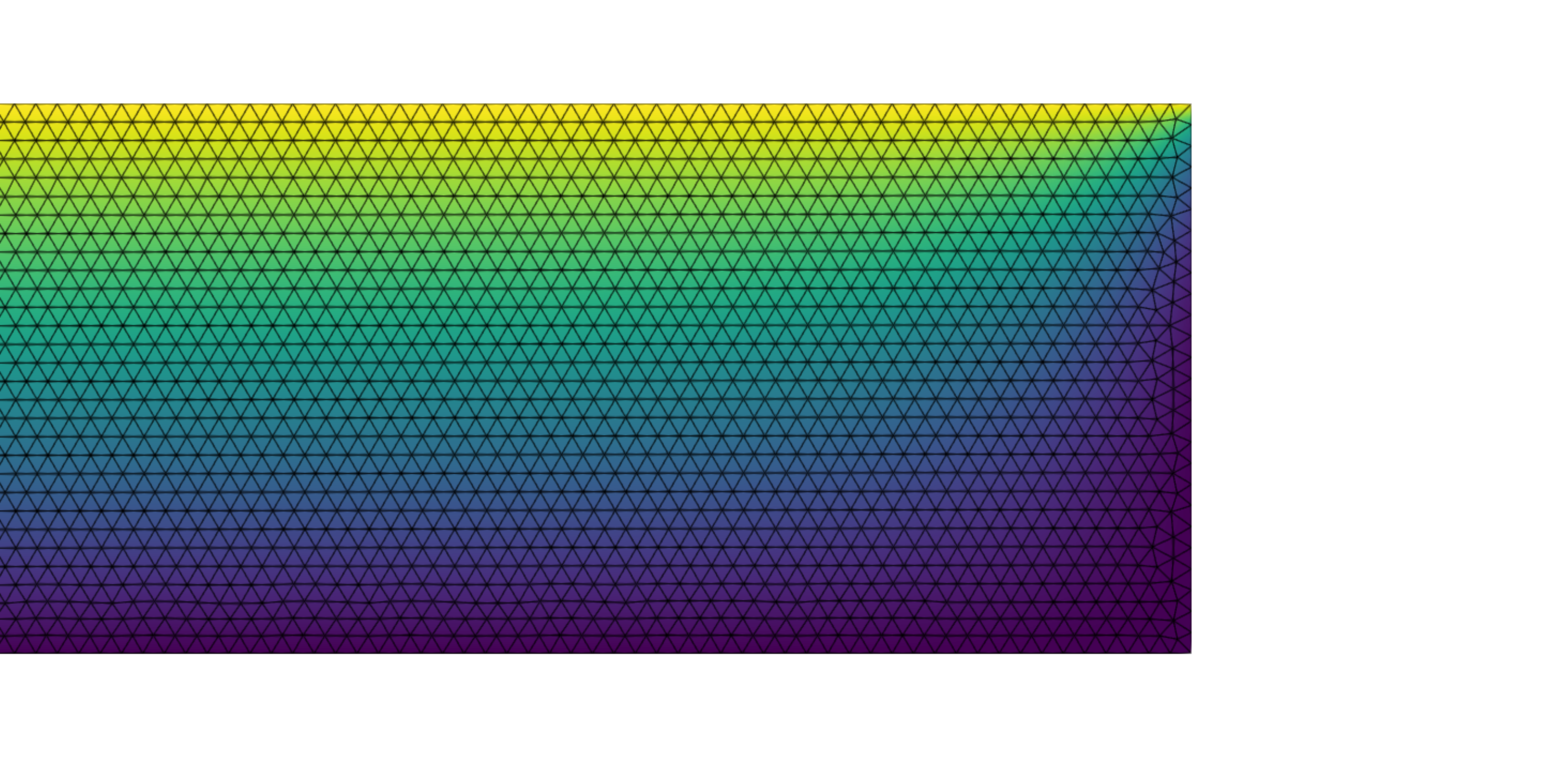}
        \caption{Salt region, uncoated}
        \label{fig:uncoated_salt}
    \end{subfigure}%
    \begin{subfigure}[b]{0.5\linewidth}
        \includegraphics[width=\linewidth,trim=0pt 650pt 0pt 450pt, clip]{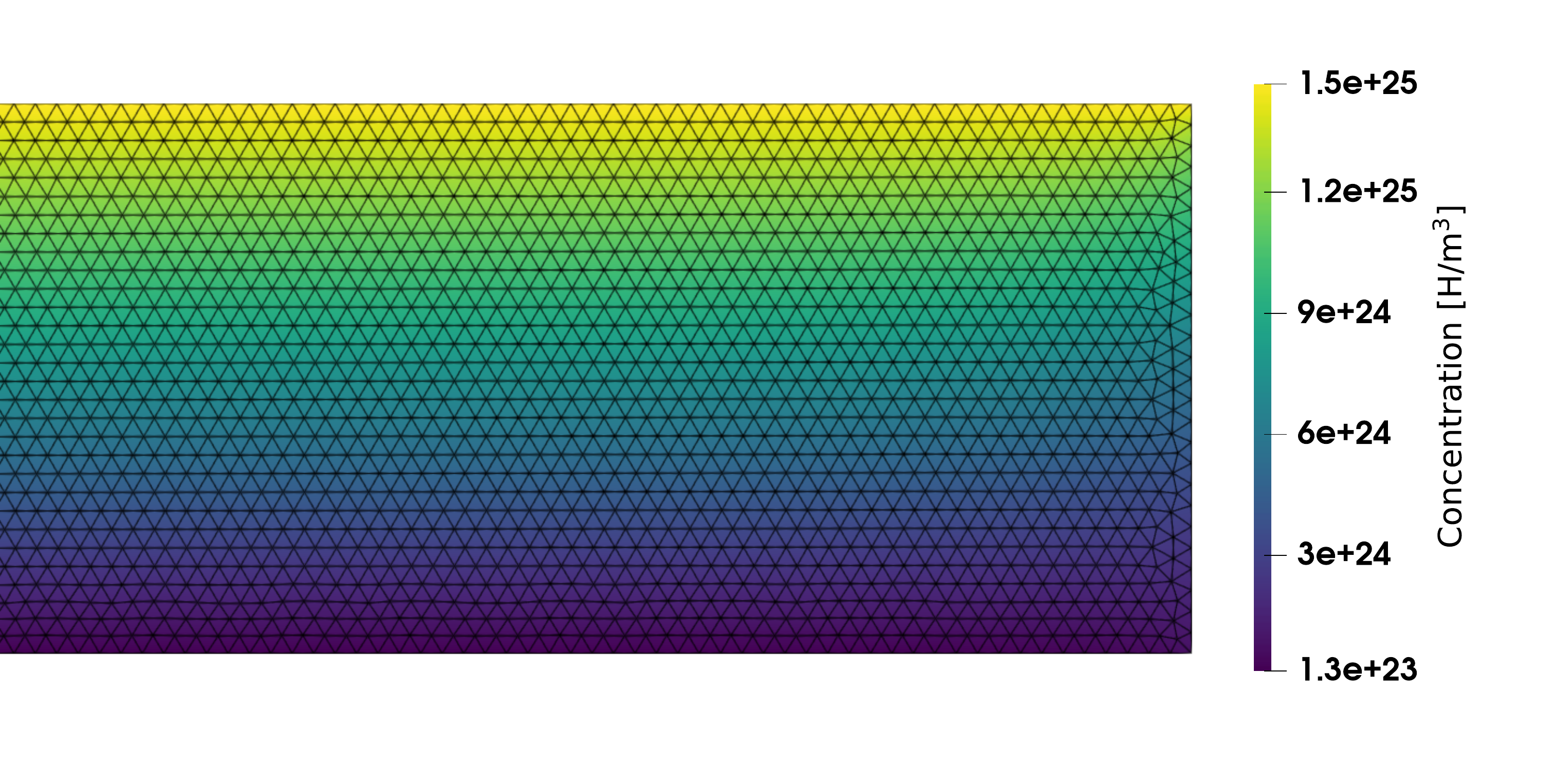}
        \caption{Salt region, ideal coating}
        \label{fig:ideal_salt}
    \end{subfigure}

    % solid
    \begin{subfigure}[b]{0.5\linewidth}
        \includegraphics[width=\linewidth, trim=0pt 0pt 0pt 650pt, clip]{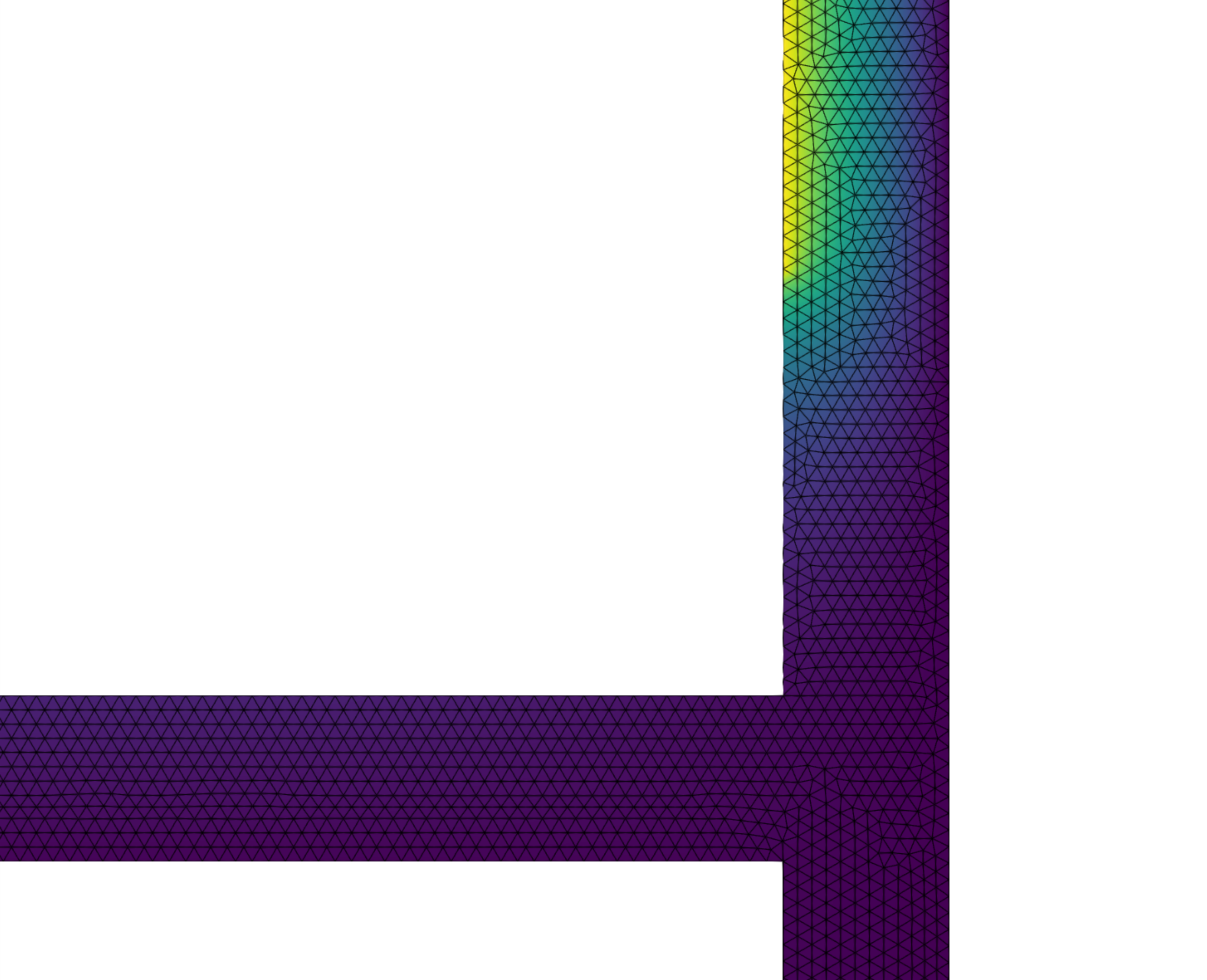}
        \caption{Solid region, uncoated}
        \label{fig:uncoated_solid}
    \end{subfigure}%
    \begin{subfigure}[b]{0.5\linewidth}
        \includegraphics[width=\linewidth, trim=0pt 0pt 0pt 650pt, clip]{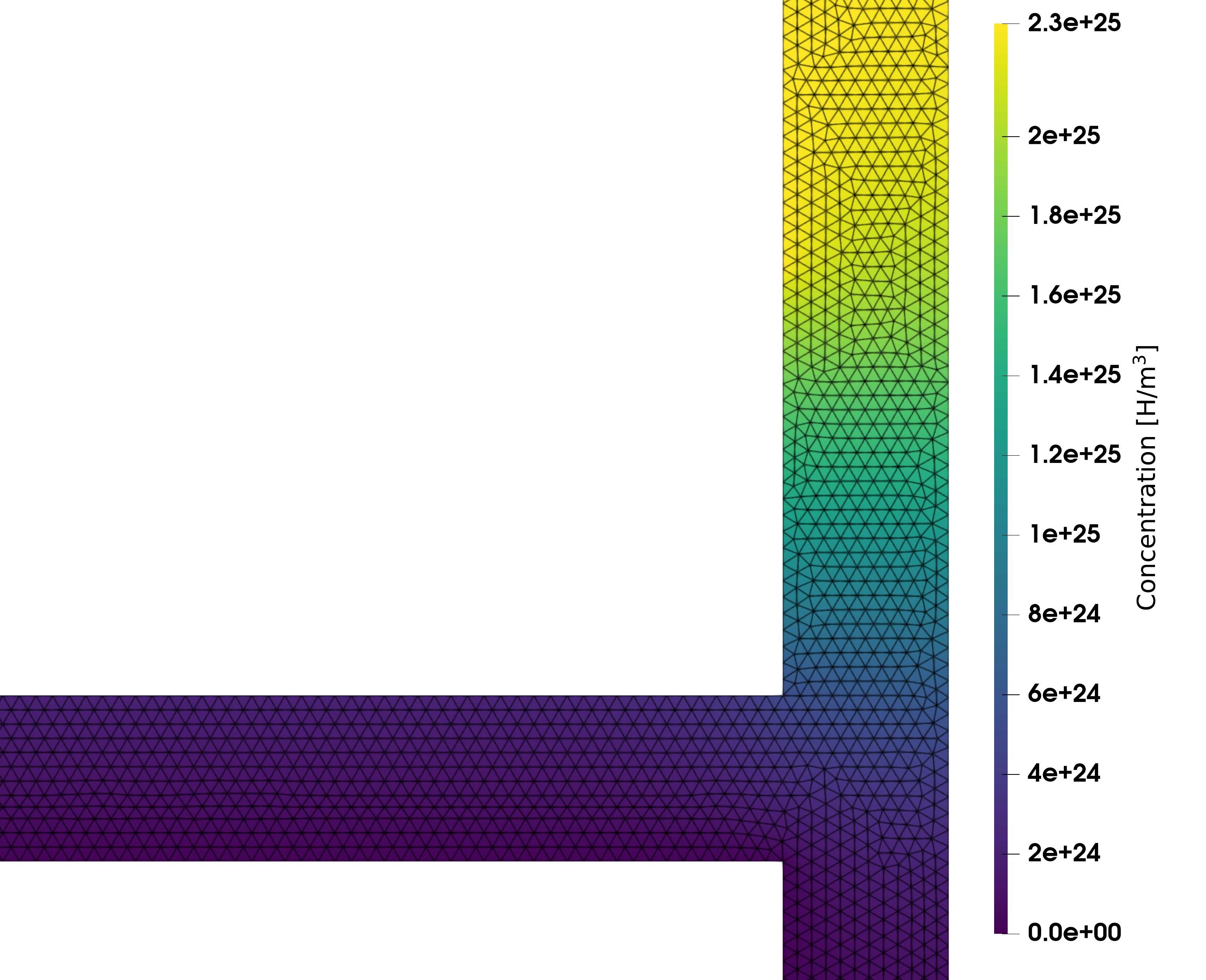}
        \caption{Solid region, ideal coating}
        \label{fig:ideal_solid}
    \end{subfigure}

    \caption{Hydrogen concentration distributions at $T = \SI{773}{\kelvin}$, with the upstream pressure of \SI{1e5}{Pa} and the downstream pressure of \SI{10}{Pa}. (a, b) Full domain with red and blue boxes indicating the magnified regions shown in (c, d) salt region and (e, f) solid region, respectively. Left column: uncoated case; right column: ideal coating case. Note that color scales differ between rows; refer to individual color bars for absolute values.}
    \label{fig:H_concentration}
\end{figure*}

The resulting hydrogen concentration distributions are shown in Fig.~\ref{fig:H_concentration}. The full-domain views (Figs.~\ref{fig:uncoated_full} and~\ref{fig:ideal_full}) are complemented by magnified views of the salt region (Figs.~\ref{fig:uncoated_salt} and~\ref{fig:ideal_salt}) and the solid region (Figs.~\ref{fig:uncoated_solid} and~\ref{fig:ideal_solid}), which reveal finer spatial details of the concentration field. Despite identical material properties, the two boundary-condition assumptions lead to distinct spatial distributions. Under the uncoated condition, hydrogen exchange with the external environment induces depletion near the outer vessel wall (Fig.~\ref{fig:uncoated_solid}) and generates lateral gradients in the salt that extend inward from the wall (Fig.~\ref{fig:uncoated_salt}). Under the ideal coating condition, outward flux is suppressed and the solid region remains close to the upstream concentration throughout (Fig.~\ref{fig:ideal_solid}), with correspondingly weaker lateral gradients in the salt.

These differences indicate that the presence of multidimensional transport pathways, including radial diffusion through the vessel structure and sidewall leakage, is absent from one-dimensional descriptions. Because the measured permeation flux integrates the response of all such pathways, the same permeability can yield different fluxes depending on the assumed external boundary, motivating the bounded reporting strategy adopted in the following sections.

\subsection{Nickel dry-run results and metal-side constraint}

Steady-state hydrogen permeation measurements in the dry-run configuration are summarized in Table~\ref{tab:dry_run_flux}. Two independent runs were conducted at each temperature to assess experimental repeatability, with uncertainties (Section~\ref{sec:UQ}) reported as one standard deviation.

\begin{table*}[h!]
\centering
\caption{Experimental hydrogen permeation measurements used for dry-run model calibration. Uncertainties are reported as one standard deviation.}
\label{tab:dry_run_flux}
\begin{tabular}{rlrrr}
\toprule
T (\si{\kelvin}) & Runs & $P_{\text{up}}$ (\si{Pa}) & $P_{\text{down}}$ (\si{Pa}) & Flux (\si{H~ s^{-1}}) \\
\midrule
\multirow{2}{*}{773}
 & Run 1 & \num{1.30e5} & \num{1.98e2} & \num{4.52(0.15)e16} \\
 & Run 2 & \num{1.10e5} & \num{1.79e2} & \num{4.08(0.14)e16} \\
\\
\multirow{2}{*}{873}
 & Run 1 & \num{1.30e5} & \num{4.59e2} & \num{1.03(0.034)e17} \\
 & Run 2 & \num{1.10e5} & \num{4.02e2} & \num{9.19(0.31)e16} \\
\\
\multirow{2}{*}{973}
 & Run 1 & \num{1.30e5} & \num{8.16e2} & \num{1.85(0.061)e17} \\
 & Run 2 & \num{1.10e5} & \num{7.36e2} & \num{1.65(0.055)e17} \\
\bottomrule
\end{tabular}
\end{table*}

\begin{figure}[htbp]
    \centering
    \includegraphics[width=\linewidth]{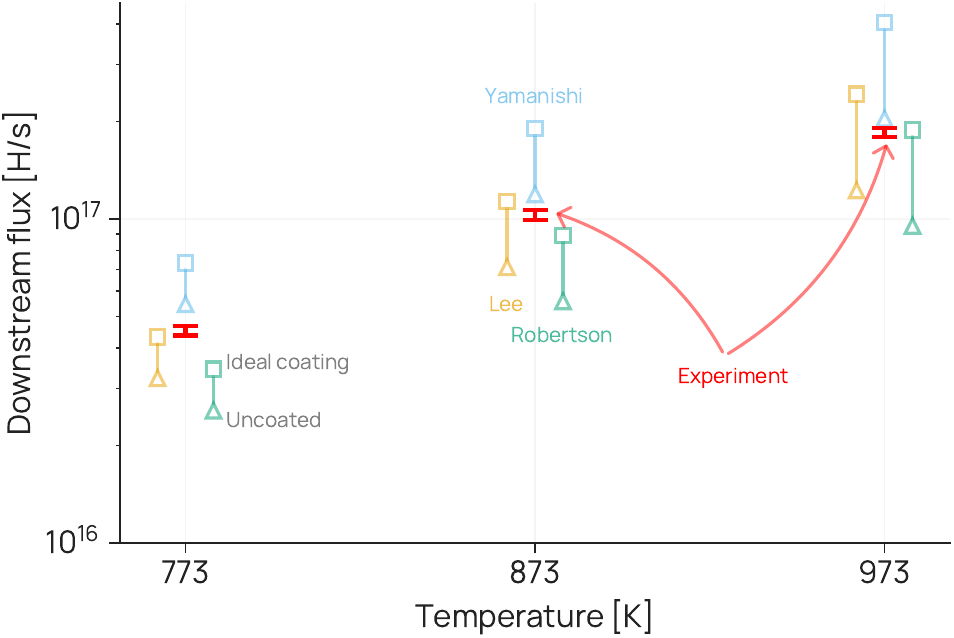}
    \caption{Comparison between experimentally measured steady-state hydrogen permeation flux and model predictions using literature nickel permeability correlations under different external boundary-condition assumptions.}
    \label{fig:dry_run_comparison}
\end{figure}

Fig.~\ref{fig:dry_run_comparison} compares the measured fluxes from Run~1 with model predictions obtained using three representative nickel permeability correlations from the literature (Yamanishi et al.~\cite{Yamanishi1983HydrogenDiffusion}, Lee et al.~\cite{Lee2013NiPermeation}, and Robertson~\cite{Robertson1973Hydrogen}). under both the ideal coating and uncoated boundary-condition limits. Run~2 yields similar trends and is omitted for clarity.

The measured fluxes are bracketed by the combined envelope of the three correlations and the two boundary-condition limits across the investigated temperature range, confirming that the multidimensional model reproduces the observed permeation behavior at the system level. 
Although no single correlation matches the data exactly, the experimental values are most consistent with the Yamanishi et al.~\cite{Yamanishi1983HydrogenDiffusion} predictions under the uncoated assumption and with the Lee et al.~\cite{Lee2013NiPermeation} and Robertson~\cite{Robertson1973Hydrogen} predictions under the ideal coating assumption. 

Across all correlations, the two boundary-condition limits produce a systematic separation: the uncoated case yields lower downstream fluxes than the ideal coating case, reflecting hydrogen losses to the surrounding environment that reduce the net flux reaching the downstream boundary.

Matching the simulated steady-state fluxes to the dry-run measurements yields the following effective nickel permeabilities for each boundary-condition limit:

\begin{equation}
% \Phi_{\mathrm{Ni}}^{\mathrm{ideal}} = 1.04\times10^{-7}
% \exp\!\left(-\frac{45.5 ~\si{kJ/mol}}{RT}\right),
\Phi^\mathrm{ideal}_\mathrm{Ni} = 
    \SI{6.26e16}
    \exp\left(-\frac{\SI{0.472}{eV}}{k_\mathrm{B}T}\right)
\end{equation}

\begin{equation}
% \Phi_{\mathrm{Ni}}^{\mathrm{uncoated}} = 9.03\times10^{-7}
% \exp\!\left(-\frac{57.6 ~\si{kJ/mol}}{RT}\right).
\Phi^\mathrm{uncoated}_\mathrm{Ni} = 
    \SI{5.44e17}
    \exp\left(-\frac{\SI{0.597}{eV}}{k_\mathrm{B}T}\right)
\end{equation}
with the unit of \si{H\,m^{-1}\,s^{-1}\,Pa^{-1/2}}.

These parameterizations are adopted in the coupled Ni-FLiBe simulations described in the following section.

\subsection{Temperature influence}
The temperature-dependent FLiBe thickness, computed from the measured salt mass and the temperature-dependent density (Section~\ref{sec:T_influence}), is summarized in Table~\ref{tab:flibe_thickness}. The thickness increases by approximately \SI{4}{\percent} over the investigated temperature range, and this variation is incorporated into all subsequent simulations.

\begin{table}[htbp]
    \centering
    \caption{Temperature-dependent FLiBe properties used to 
    determine the effective salt thickness in the HYPERION 
    configuration.}
    \label{tab:flibe_thickness}
    \begin{tabular}{cc}
        \hline
        Temperature (\si{\kelvin}) & Salt thickness (\si{mm}) \\
        \hline
        773 & 5.140 \\
        823 & 5.194 \\
        873 & 5.249 \\
        923 & 5.306 \\
        973 & 5.364 \\
        \hline
    \end{tabular}
\end{table}

\subsection{Inferred FLiBe permeability}
\label{sec:flibe_results}

Coupled Ni-FLiBe simulations are performed at each experimental temperature to infer the FLiBe permeability of hydrogen and deuterium. The measured downstream fluxes are summarized in Tables~\ref{tab:run2_flux} and \ref{tab:run3_flux}.

\begin{table*}[h!]
\centering
\caption{Hyperion experiment summary for H$_2$ permeation (steady state). Uncertainties are reported as one standard deviation.}
\label{tab:run2_flux}
\begin{tabular}{c c c c}
\toprule
T (\si{\kelvin}) & $P_{\text{up}}$ (Pa) & $P_{\text{down}}$ (Pa) & Flux (H/s) \\
\midrule
773 & \num{1.31e5} & \num{1.99e1} & \num{4.34(0.15)e15} \\
823 & \num{1.31e5} & \num{3.89e1} & \num{8.58(0.29)e15} \\
873 & \num{1.32e5} & \num{4.62e1} & \num{1.01(0.034)e16} \\
923 & \num{1.32e5} & \num{5.02e1} & \num{1.10(0.036)e16} \\
973 & \num{1.32e5} & \num{4.78e1} & \num{1.04(0.035)e16} \\
\bottomrule
\end{tabular}
\end{table*}

\begin{table*}[h!]
\centering
\caption{Hyperion experiment summary for D$_2$ permeation (steady state). Uncertainties are reported as one standard deviation.}
\label{tab:run3_flux}
\begin{tabular}{c c c c}
\toprule
T (\si{\kelvin}) & $P_{\text{up}}$ (Pa) & $P_{\text{down}}$ (Pa) & Flux (D/s) \\
\midrule
773 & \num{1.31e5} & \num{8.66}   & \num{1.91(0.065)e15} \\
873 & \num{1.33e5} & \num{2.10e1} & \num{4.50(0.15)e15} \\
973 & \num{1.31e5} & \num{3.23e1} & \num{7.12(0.24)e15} \\
\bottomrule
\end{tabular}
\end{table*}

\begin{figure}[h!]
    \centering
    \includegraphics[width=\linewidth]{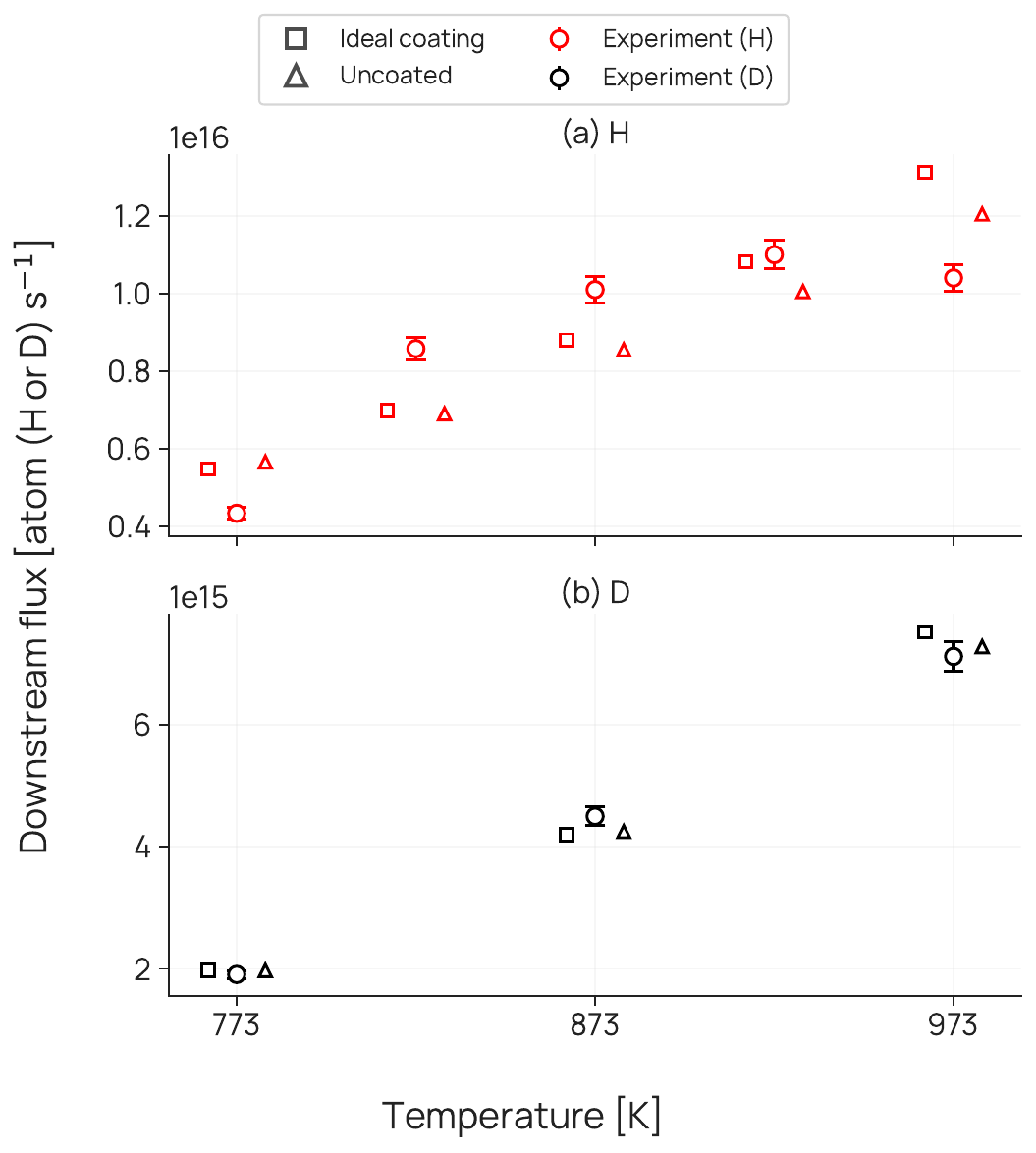}
    \caption{Comparison between simulated and experimentally measured steady-state hydrogen permeation flux under different external boundary condition assumptions.(a) for H and (b) for D.}
    \label{fig:BC_comparison}
\end{figure}

The simulated and measured fluxes are compared in Fig.~\ref{fig:BC_comparison} for both isotopes and both boundary-condition limits. 

The inferred FLiBe permeabilities follow Arrhenius form:
For hydrogen:
\begin{align}
\Phi^\mathrm{ideal}_\mathrm{H}    &= 8.17 \times 10^{11} \exp\left(-\frac{0.206~\si{eV}}{k_\mathrm{B}T}\right), \\
\Phi^\mathrm{uncoated}_\mathrm{H} &= 4.23 \times 10^{13} \exp\left(-\frac{0.429~\si{eV}}{k_\mathrm{B}T}\right),
\end{align}
with units of \si{H\,m^{-1}\,s^{-1}\,Pa^{-1}}.

For deuterium:
\begin{align}
\Phi^\mathrm{ideal}_\mathrm{D}    &= 2.20 \times 10^{12} \exp\left(-\frac{0.381~\si{eV}}{k_\mathrm{B}T}\right), \\
\Phi^\mathrm{uncoated}_\mathrm{D} &= 2.85 \times 10^{14} \exp\left(-\frac{0.615~\si{eV}}{k_\mathrm{B}T}\right),
\end{align}
with units of \si{D\,m^{-1}\,s^{-1}\,Pa^{-1}}.

% \begin{equation}
% % \Phi_{\mathrm{H}}^{\mathrm{ideal}}= 1.36\times10^{-12}
% % \exp\left(-\frac{19.92~\si{kJ/mol}}{RT}\right),
% \Phi^\mathrm{ideal}_\mathrm{H} = 
%     \SI{8.17e11}
%     \exp\left(-\frac{\SI{0.206}{eV}}{k_\mathrm{B}T}\right)
% \end{equation}

% \begin{equation}
% % \Phi_{\mathrm{H}}^{\mathrm{uncoated}} = 7.04\times10^{-11}
% % \exp\left(-\frac{41.37~\si{kJ/mol}}{RT}\right),
% \Phi^\mathrm{uncoated}_\mathrm{H} = 
%     \SI{4.23e13}
%     \exp\left(-\frac{\SI{0.429}{eV}}{k_\mathrm{B}T}\right)
% \end{equation}
% with the unit of \si{H\,m^{-1}\,s^{-1}\,Pa^{-1}}

% \begin{equation}
% % \Phi_{\mathrm{D}}^{\mathrm{ideal}} = 3.64\times10^{-12}
% % \exp\left(-\frac{36.71~\si{kJ/mol}}{RT}\right),
% \Phi^\mathrm{ideal}_\mathrm{D} = 
%     \SI{2.20e12}
%     \exp\left(-\frac{\SI{0.381}{eV}}{k_\mathrm{B}T}\right)
% \end{equation}

% \begin{equation}
% % \Phi_{\mathrm{D}}^{\mathrm{uncoated}} = 4.73\times10^{-10}
% % \exp\left(-\frac{59.35~\si{kJ/mol}}{RT}\right).
% \Phi^\mathrm{uncoated}_\mathrm{D} = 
%     \SI{2.85e14}
%     \exp\left(-\frac{\SI{0.615}{eV}}{k_\mathrm{B}T}\right)
% \end{equation}
% with the unit of \si{D\,m^{-1}\,s^{-1}\,Pa^{-1}}

\begin{figure}[h]
    \centering
    \includegraphics[width=\linewidth]{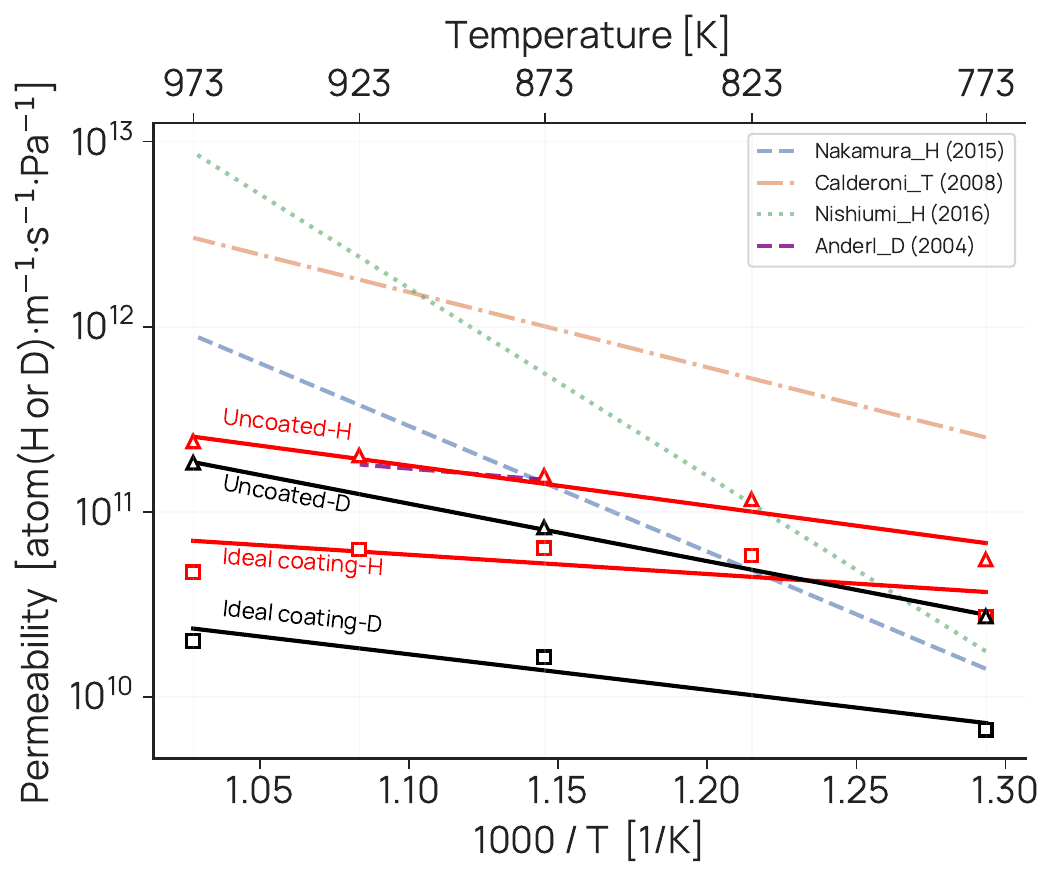}
    \caption{Arrhenius representation of the inferred FLiBe permeability for hydrogen and deuterium under the two external boundary-condition assumptions, compared with literature correlations.}
    \label{fig:flibe_permeability}
\end{figure}

Fig.~\ref{fig:flibe_permeability} compares these results with representative literature correlations.

The permeability inferred under the uncoated condition is systematically higher by more than an order of magnitude than that obtained under the ideal coating assumption. The inferred values are generally lower than most literature correlations reported for FLiBe, with the discrepancy being more pronounced at higher temperatures.

%% file: discussion.tex
\section{Discussion}
\label{sec:discussions}

\subsection{Multidimensional transport pathways and their quantitative role}
\label{sec:sidewall_transport}

Beyond the primary axial permeation route, the HYPERION geometry includes nickel sidewall surfaces that introduce additional transport pathways, as illustrated in Fig.~\ref{fig:flow_path}. The role of these pathways is quantified below using the inferred FLiBe permeability under both boundary-condition assumptions.

\begin{figure}[h!]
    \centering
    \includegraphics[width=\linewidth]{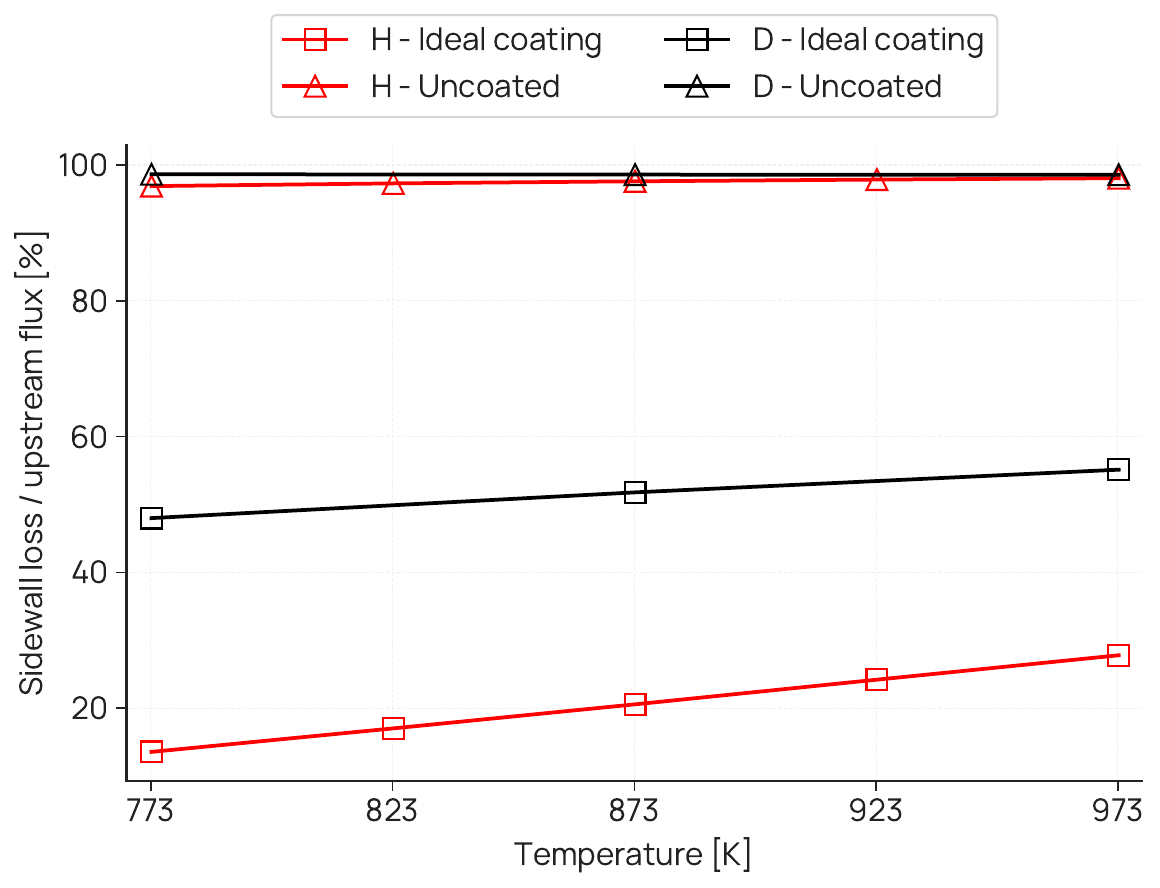}
    \caption{Sidewall flux loss normalized by upstream permeation flux.}
    \label{fig:sidewall_loss_up}
\end{figure}

Fig.~\ref{fig:sidewall_loss_up} shows the fraction of upstream hydrogen and deuterium flux lost through the sidewalls, defined as
\begin{equation}
    f_\mathrm{loss} = \left(1 - \frac{J_\mathrm{liquid}}{J_\mathrm{in}}\right)\times 100\%,
    \label{eq:f_leak}
\end{equation}
where $J_\mathrm{in}$ is the total flux integrated over all upstream surfaces, and $J_\mathrm{liquid}$ is the flux entering through the 
free liquid surface, representing the primary intended permeation pathway. Under uncoated conditions, nearly the entire upstream flux (~\SI{97}{\percent} for both isotopes) is diverted through the vessel structure, leaving only a small fraction following the intended axial pathway. Under ideal-coating conditions, sidewall losses are much smaller and increase modestly with temperature: from \SI{14}{\percent} to \SI{28}{\percent} for hydrogen and from \SI{48}{\percent} to \SI{55}{\percent} for deuterium across the temperature range. Notably, the deuterium sidewall loss under ideal coating is consistently higher than that of hydrogen, by roughly a factor of two to three across the temperature range.

\begin{figure}[h!]
    \centering
    \includegraphics[width=\linewidth]{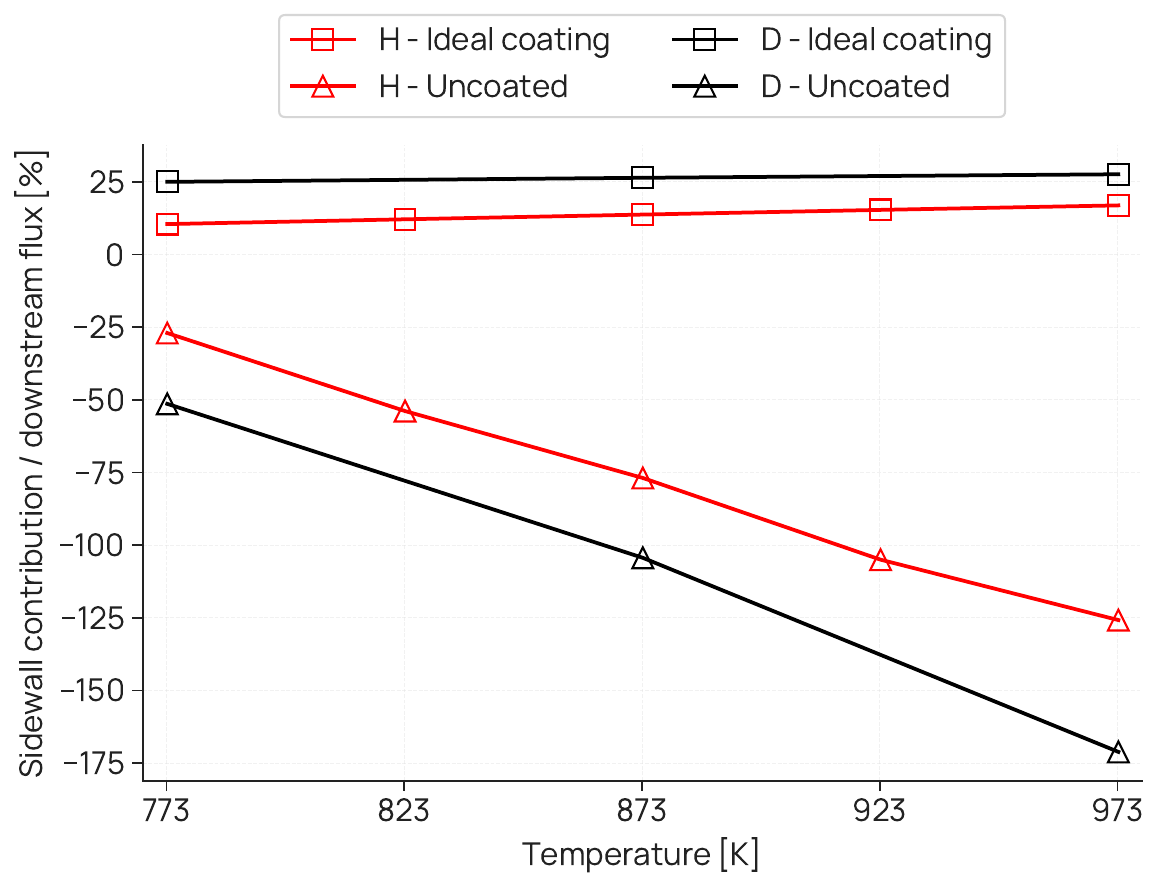}
    \caption{Sidewall flux contribution normalized by downstream permeation flux.}
    \label{fig:sidewall_contribution_down}
\end{figure}

Fig.~\ref{fig:sidewall_contribution_down} presents the net contribution of sidewall transport to the downstream flux, defined as
\begin{equation}
    f_\mathrm{contribution} = \left(1 - \frac{J_\mathrm{membrane}}{J_\mathrm{out}}\right)\times100\%,
    \label{eq:f_sidewall}
\end{equation}
where $J_\mathrm{membrane}$ is the flux through the Ni membrane into the downstream sweep gas, representing the primary intended permeation pathway; $J_\mathrm{out}$ is the net flux summed over all downstream side surfaces of the computational domain, including the membrane and the surrounding downstream sidewalls. With outward flux from the domain taken as positive, a positive $f_\mathrm{contribution}$ indicates that the sidewalls act as a bypass that augments the measured downstream signal, while a negative value indicates that the sidewalls serve as a net hydrogen sink, with hydrogen leaving the downstream region laterally. The sidewall outflow is taken to discharge directly into the glovebox environment. A magnitude exceeding \SI{100}{\percent} indicates that this sidewall loss exceeds the total flux reaching the downstream detector.

Under ideal coating, \SIrange{10}{17}{\percent} of the hydrogen and \SIrange{25}{27}{\percent} of the deuterium detected downstream originates from sidewall-mediated bypass through the vessel structure rather than direct transport through the Ni membrane. Under uncoated conditions, $f_\mathrm{contribution}$ becomes strongly negative and grows in magnitude with temperature, from \SI{-27}{\percent} at \SI{773}{\kelvin} to \SI{-126}{\percent} at \SI{973}{\kelvin} for hydrogen, and from \SI{-51}{\percent} to \SI{-171}{\percent} for deuterium. At the highest temperatures, more hydrogen is therefore lost laterally to the glovebox than is detected at the downstream port.

Taken together, these results show that the sidewalls play a dual role depending on the external boundary condition: under ideal coating they provide supplementary bypass pathways that augment the downstream flux, whereas under uncoated conditions, they act as a dominant loss mechanism to the surrounding environment. In both cases, the magnitude of sidewall transport is comparable to or larger than the primary membrane flux, indicating that one-dimensional representations cannot capture the system response.

\subsection{Bias in permeability inference due to multidimensional transport}
\label{sec:bias}

\begin{figure}[htbp]
    \centering
    \includegraphics[width=\linewidth]{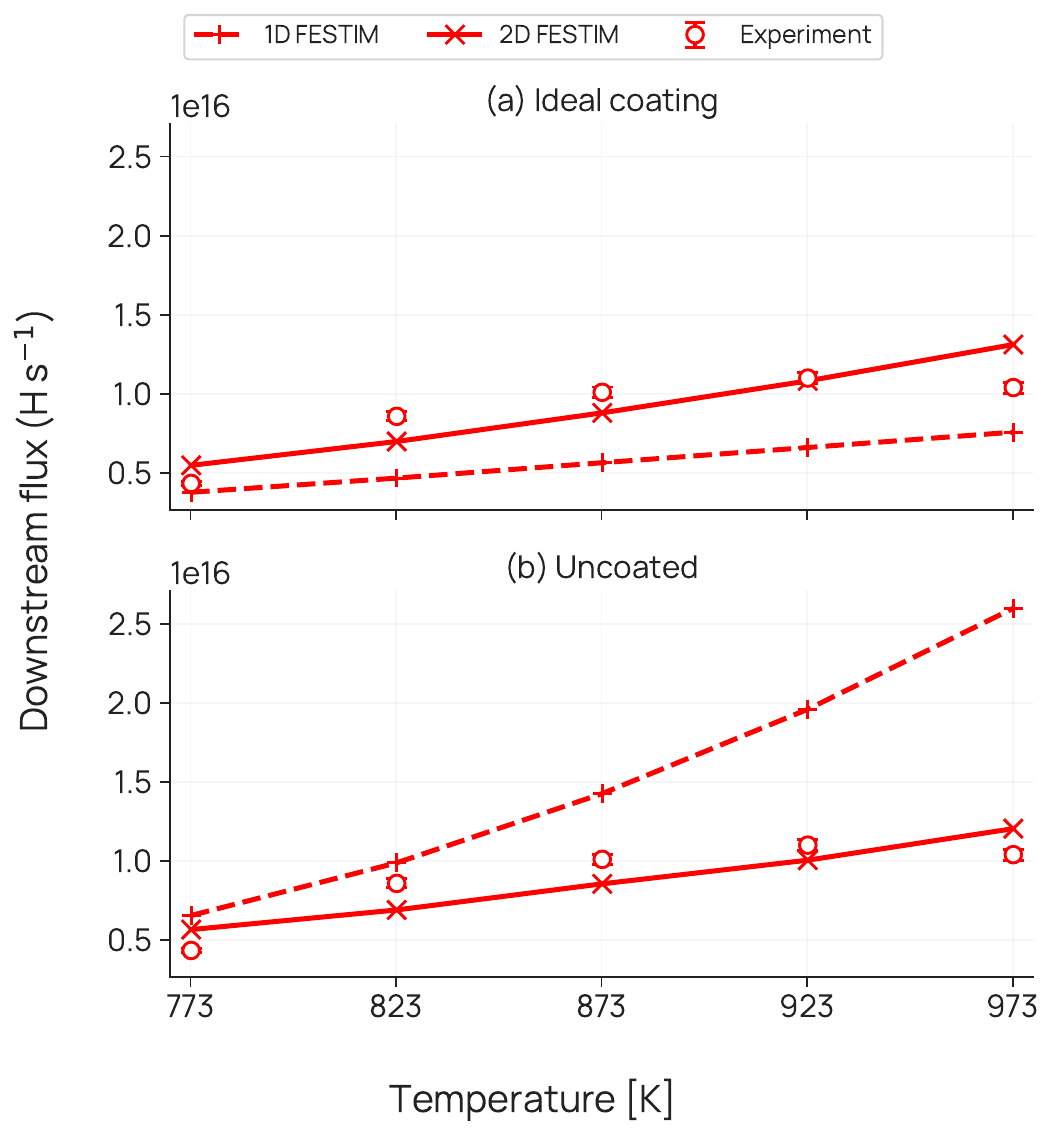}
    \caption{Comparison of downstream hydrogen flux predicted by one-dimensional (1D) and two-dimensional (2D) FESTIM models with experimental measurements under (a) ideal-coating and (b) uncoated boundary conditions.}
    \label{fig:1D_2D_comparison_H}
\end{figure}

\begin{figure}[htbp]
    \centering
    \includegraphics[width=\linewidth]{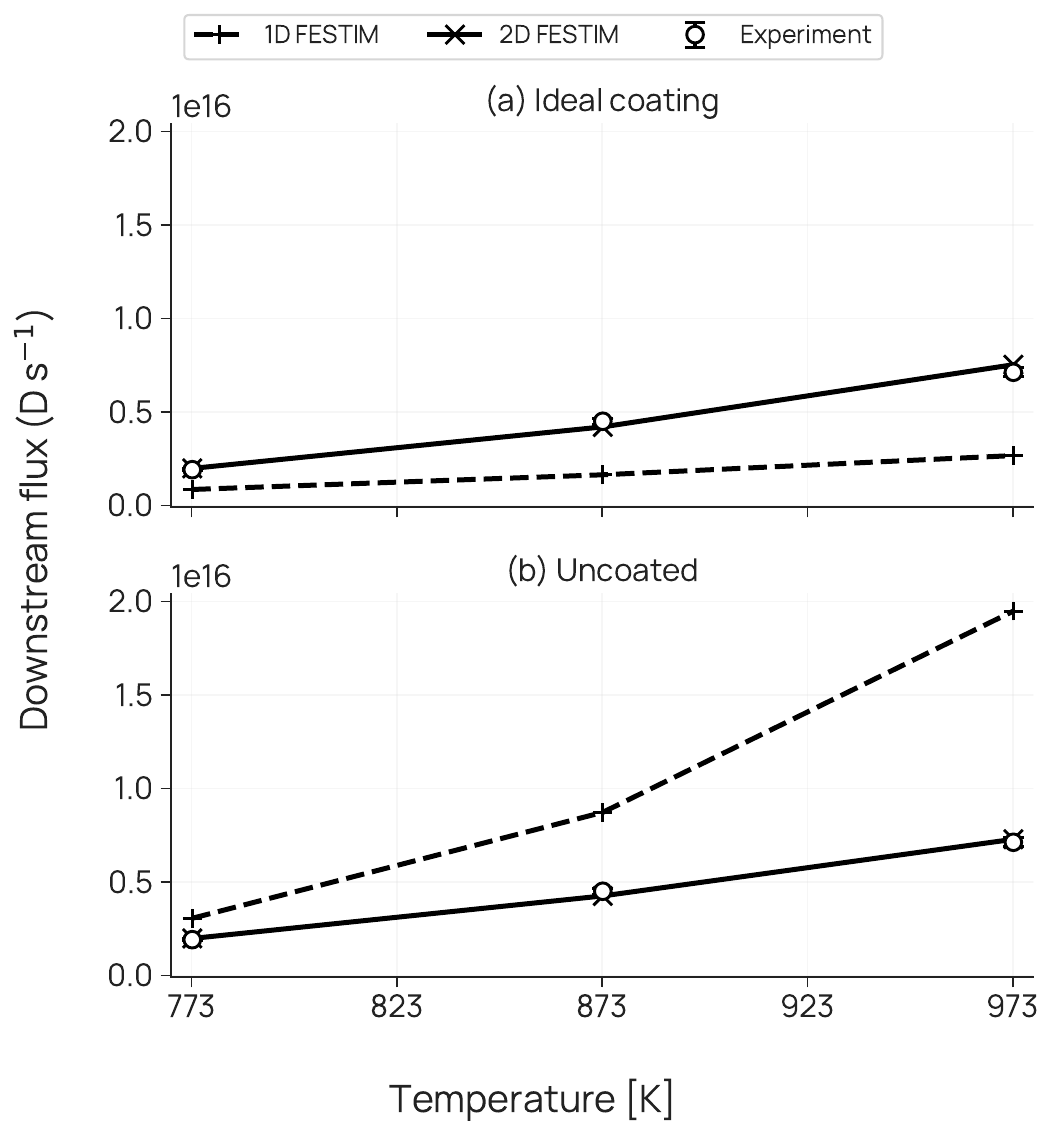}
    \caption{Comparison of downstream deuterium flux predicted by one-dimensional (1D) and two-dimensional (2D) FESTIM models with experimental measurements under (a) ideal-coating and (b) uncoated boundary conditions.}
    \label{fig:1D_2D_comparison_D}
\end{figure}

The quantitative significance of sidewall transport established in Section~\ref{sec:sidewall_transport} has direct consequences for permeability inference. To evaluate these consequences, predictions from one-dimensional (1D) and two-dimensional (2D) FESTIM models are compared against experimental measurements for hydrogen and deuterium using the corresponding inferred permeability, as shown in Fig.~\ref{fig:1D_2D_comparison_H} and Fig.~\ref{fig:1D_2D_comparison_D}. Both models use the same FLiBe permeability obtained from the 2D inverse procedure (Section~\ref{sec:flibe_results}). The 2D model therefore reproduces the experimental measurements by construction, while the 1D predictions reveal what would be observed if the same material property were used in a 1D formulation. Equivalently, how much the 1D inverse procedure would need to adjust its inferred permeability to match the data.

Under ideal-coating boundary conditions (Fig.~\ref{fig:1D_2D_comparison_H}a and \ref{fig:1D_2D_comparison_D}a), the 1D model systematically underpredicts the downstream flux for both hydrogen and deuterium across the full temperature range, whereas the 2D model provides significantly improved agreement with the experimental measurements. This discrepancy reflects the bypass pathways through the vessel structure that, under ideal coating, contribute \SIrange{10}{17}{\percent} of the downstream hydrogen flux and \SIrange{25}{27}{\percent} of the deuterium flux (Section~\ref{sec:sidewall_transport}) and are not captured by the 1D formulation. If the 1D model were instead used to infer permeability from the measured fluxes, as is common in the literature, the fitted values would have to be artificially increased to compensate for the underprediction, yielding a systematic overestimation of permeability. This is consistent with the observation that the FLiBe permeabilities inferred in the present study lie below most literature correlations (Fig.~\ref{fig:flibe_permeability}).

Under uncoated boundary conditions (Figs.~\ref{fig:1D_2D_comparison_H}b and~\ref{fig:1D_2D_comparison_D}b), the 1D model exhibits the opposite behavior: it overestimates the downstream flux, with the discrepancy growing strongly with temperature and reaching approximately a factor of 2.5 at \SI{973}{\kelvin} for both isotopes. This is consistent with the strongly negative sidewall contribution under uncoated conditions identified in Fig.~\ref{fig:sidewall_contribution_down} (\SIrange{-27}{-126}{\percent} for hydrogen and \SIrange{-51}{-171}{\percent} for deuterium): hydrogen entering the sidewall structure escapes to the glovebox rather than reaching the downstream detector, reducing the measured flux below what axial transport alone would produce. Because the 1D model lacks this loss mechanism, the bias grows with temperature as sidewall losses become increasingly dominant.

Taken together, these results show that the 1D formulation produces opposite-sign errors depending on the external boundary condition: it underpredicts the downstream flux under ideal coating and overpredicts it under uncoated conditions. The direction and magnitude of the bias depend on both the boundary condition and the operating temperature, making the inferred permeability from a 1D analysis unreliable in coupled systems where structural transport pathways are present.

The key question is not whether a 1D model can be made to reproduce the measured flux for a specific experiment-By adjusting permeability, it generally can-but whether the resulting permeability is a transferable material property. A value that implicitly absorbs sidewall losses or bypass contributions inherits the geometry and external boundary conditions of the experiment in which it was inferred, and cannot be used predictively in a different system, such as a reactor-scale blanket simulation. The multidimensional framework, by resolving these pathways explicitly, separates geometric and boundary-condition contributions from the material response, yielding a permeability conditioned on a physically consistent representation of the system rather than on an idealized 1D abstraction.

\subsection{Implications for FLiBe permeability data in the literature}
\label{sec:implications}

The permeability values inferred in this work lie below most literature correlations reported for FLiBe, particularly at higher temperatures (Fig.~\ref{fig:flibe_permeability}).

Before discussing this trend, a specific feature of the hydrogen data warrants attention: the measured flux at \SI{973}{\kelvin} is slightly lower than at \SI{923}{\kelvin}, deviating from a monotonic temperature dependence. This behavior may be associated with interfacial effects at elevated temperatures, such as gas bubble formation at the liquid-solid interface reported in the experimental work~\cite{Saraswat2026FLiBePermeation}, which could reduce the effective interfacial area and introduce additional mass-transfer resistance not captured by the present model.

Beyond this specific high-temperature effect, the present results span a range that overlaps with the lower portion of reported FLiBe permeability correlations, with the ideal-coating limit lying below most literature values and the uncoated limit approaching those reported by Anderl et al.~\cite{Anderl2004Flibe}. Rather than identifying any single literature value as biased, this comparison illustrates that the inferred permeability depends sensitively on how multidimensional transport pathways and external boundary conditions are represented in the analysis. The same underlying salt transport behavior, when interpreted under different geometric or boundary-condition assumptions, can yield permeability values differing by more than an order of magnitude, as illustrated by the spread between our two bounding cases.

Existing FLiBe permeability data have been obtained under a range of experimental geometries and analyzed predominantly within 1D frameworks. Without geometry-specific multidimensional analysis of each configuration, it is not possible to determine whether individual literature values would shift upward or downward if reinterpreted with explicit treatment of structural transport pathways: the direction depends on whether sidewall transport in that specific configuration acts as a bypass to the downstream region or as a loss to the surroundings (Sections~\ref{sec:sidewall_transport} and~\ref{sec:bias}).

The present work therefore does not claim that literature values are systematically biased in a particular direction. Instead, the scatter of approximately one to two orders of magnitude observed across the literature is consistent in magnitude with the multidimensional and boundary-condition effects quantified here, suggesting that part of this scatter may originate from differences in experimental geometry rather than from intrinsic material variability alone. Other well-established factors, including salt redox state, impurity content, and surface conditions~\cite{Lam2021HydrogenValence, Tijssen2025Permeation}, remain important contributors to the observed variability and are not addressed by the present analysis. Reliable extraction of intrinsic permeability from permeation experiments in coupled systems therefore requires explicit treatment of structural transport pathways and external boundary conditions within a multidimensional modeling framework.

\subsection{Limitations and outlook}
\label{sec:limitations}

Several sources of uncertainty remain in the present analysis, arising from both experimental measurements and modeling assumptions.

The most significant limitation concerns interfacial bubble formation at the Ni-FLiBe interface. Previous work~\cite{Saraswat2026FLiBePermeation} demonstrated that hydrogen charging at the metal side leads to bubble nucleation and growth at the liquid-solid interface, suppressing the effective permeability by up to \SI{77}{\percent} relative to salt-side charging configurations. This bubble-induced transport barrier is further supported by an observed isotope masking effect under metal-side charging. The present model assumes ideal series coupling at the liquid-solid interface, with continuous thermodynamic equilibrium and no reduction in effective interfacial area, and therefore does not capture bubble nucleation, growth, or the associated interfacial resistance. This limitation is particularly relevant above \SI{873}{\kelvin}, where bubble effects persist even under salt-side charging~\cite{Saraswat2026FLiBePermeation}, and may contribute to the deviation from monotonic temperature dependence observed in the hydrogen flux at \SI{973}{\kelvin}. Incorporating interfacial bubble dynamics, including nucleation thresholds, growth kinetics, and their effect on effective interfacial area, is a key direction for future model development.

A second source of uncertainty is the external boundary condition at the glovebox surface. Although the glovebox hydrogen partial pressure $P_\mathrm{gb}$ was monitored in real time during the HYPERION experiments using a calibrated electrochemical sensor~\cite{Saraswat2026FLiBePermeation}, the measured values cannot serve as a quantitative constraint on the sidewall flux: periodic argon purging of the glovebox removes a fraction of the leaked hydrogen before the mass balance can be closed. The two limiting boundary conditions (ideal coating and uncoated vessel) are therefore retained to bracket the uncertainty. Reducing this uncertainty in future experiments would require either a sealed glovebox configuration that permits full mass balance accounting, or direct instrumentation of the sidewall flux itself.

More broadly, these limitations indicate that future permeation experiments aimed at directly measuring FLiBe permeability should be conceived from the outset in close coordination with multidimensional modeling, rather than relying on post-hoc corrections to a 1D framework. Beyond the geometric and instrumentation improvements discussed above, diagnostics capable of characterizing the effective interfacial area under operating conditions would be needed to constrain bubble effects. Together, these developments would enable more reliable extraction of intrinsic FLiBe transport properties and support the design of next-generation permeation experiments with reduced systematic uncertainty.

%% file: conclusions.tex
\section{Conclusions}
\label{sec:conclusions}

This work demonstrates that the permeation flux measured in coupled gas-liquid-solid systems depends not only on the transport properties of molten FLiBe and the operating conditions, but also on the geometry of the experimental configuration and the external boundary conditions. When such system-level fluxes are interpreted using simplified transport models that do not account for these geometric and boundary-condition effects, the inferred permeability reflects the coupled response of the entire system rather than the salt transport properties alone.

Using a multidimensional, multi-material transport framework implemented in FESTIM, hydrogen and deuterium permeation measurements from the HYPERION Ni-FLiBe system were analyzed over the temperature range \SIrange{773}{973}{\kelvin}. The principal findings are:

\begin{itemize}

\item Sidewall transport through the nickel containment structure constitutes a significant fraction of the overall hydrogen flux. Under ideal coating, sidewall bypass pathways augment the downstream flux by \SIrange{10}{17}{\percent} for hydrogen and \SIrange{25}{27}{\percent} for deuterium. Under uncoated conditions, sidewall losses to the glovebox dominate and can exceed the flux reaching the downstream detector, by up to a factor of $\sim$2.3 for hydrogen and $\sim$2.7 for deuterium at \SI{973}{\kelvin}.

\item The inferred FLiBe permeability spans more than an order of magnitude between the ideal-coating and uncoated limits for both isotopes, reflecting the sensitivity of permeability inference to boundary-condition assumptions rather than uncertainty in the material property itself.

\item One-dimensional transport models produce qualitatively opposite errors depending on the external boundary condition: the downstream flux is systematically underpredicted under ideal coating, due to neglected sidewall bypass, and overpredicted by approximately a factor of 2.5 at \SI{973}{\kelvin} under uncoated conditions, due to neglected sidewall losses. In neither case does the 1D model represent the underlying transport physics, making 1D-based permeability inference unreliable in coupled systems where structural transport pathways are present.

\item The inferred FLiBe permeability lies in the lower portion of reported literature correlations. The present work does not establish a single corrected value, but rather shows that geometry- and boundary-condition-induced uncertainty can account for a substantial portion of the scatter in reported literature data, with implications for the transferability of any 1D-inferred permeability across different experimental configurations.

\end{itemize}

More broadly, this study establishes a validation-informed inverse framework for permeability inference in coupled molten-salt systems, in which multidimensional transport pathways and external boundary conditions are explicitly resolved. These findings have direct implications for tritium transport modeling in fusion blanket systems and underscore the importance of multidimensional approaches in the design and interpretation of permeation experiments for liquid breeder materials. As multidimensional modeling capabilities mature, future permeation experiments and their analysis frameworks should be co-designed, enabling more reliable extraction of intrinsic transport properties and direct application of permeation data to reactor-scale tritium transport modeling.